\documentclass[11pt,a4paper]{article}
\pdfoutput=1
\usepackage{jheppub}

\usepackage{amsmath}
\usepackage{verbatim}
\usepackage{amssymb}
\usepackage{inputenc, array}
\usepackage{textcomp}
\usepackage{bm}
\usepackage{physics}
\usepackage{simpler-wick}
\usepackage{appendix}
\usepackage{xcolor}

\newcommand{\e}{\epsilon}
\newcommand{\be}[1]{\begin{equation}\label{#1} }
\newcommand{\ee}{\end{equation}}
\newcommand{\bea}[1]{\begin{eqnarray}\label{#1} }
\newcommand{\eea}{\end{eqnarray}}
\newcommand{\refb}[1]{(\ref{#1})}

\newcommand{\z}{{\bar z}}
\newcommand{\h}{{\bar h}}

\renewcommand{\>}{\rangle}
\newcommand{\<}{\langle}

\renewcommand{\t}{\tau}

 \newcommand{\p}{\textbf{p}}
\newcommand{\x}{\textbf{x}}
\newcommand{\y}{\textbf{y}}
\pdfoutput=1
\usepackage{array}
\usepackage{multirow}
\usepackage{amsmath}
\usepackage{makecell}

\definecolor{green}{rgb}{0.1,0.8,0.2}

\usepackage{cancel}
\usepackage{graphicx} 

\usepackage{hyperref}
\usepackage{cleveref}
\usepackage{float}
\usepackage{subcaption}
\usepackage{physics}
\usepackage{slashed}
\usepackage{amsfonts}
\usepackage{caption}
\usepackage{mathrsfs}

\allowdisplaybreaks
\title{AdS Witten Diagrams to Carrollian Correlators}

\author[a, b]{Arjun Bagchi,}  \author[a]{Prateksh Dhivakar,} \author[a]{and Sudipta Dutta.}\author{\\}

\affiliation[a]{Indian Institute of Technology Kanpur, Kanpur 208016, India.\\} 

\affiliation[b]{Centre de Physique Theorique, Ecole Polytechnique de Paris, 91128 Palaiseau Cedex, France.\\}

\emailAdd{(abagchi, prateksh, dsudipta)@iitk.ac.in}

\preprint{}
\abstract{Carrollian Conformal Field Theories (CFTs) have been proposed as co-dimension one holographic duals to asymptotically flat spacetimes as opposed to Celestial CFTs which are co-dimension two. In this paper, drawing inspiration from Celestial holography, we show by a suitable generalisation of the flat space limit of AdS that keeps track of the previously disregarded null direction, one can reproduce Carrollian CFT correlation functions from AdS Witten diagrams. In particular, considering Witten diagrams in AdS$_4$, we reproduce two and three-point correlation functions for three dimensional Carrollian CFTs in the so-called delta-function branch. Along the way, we construct non-trivial Carrollian three-point functions in the delta-branch by considering a collinear limit. We also obtain a generalised anti-podal matching condition that now depends on the retarded time direction.}

\begin{document}
\maketitle

\vfill
\newpage
\section{Introduction}
The Holographic Principle \cite{tHooft:1993dmi, Susskind:1994vu}, and in particular its incarnation in Anti de Sitter spacetimes via the AdS/CFT correspondence \cite{Maldacena:1997re,Witten:1998qj}, over the past decades has been our most useful tool to unravel the mysteries associated with quantum gravity. The progress in understanding holography beyond its natural setting in AdS has been somewhat less impressive. Of late, there is a renewed effort in the construction of holography for asymptotically flat spacetimes. 

\subsection*{Two approaches to Flat Holography}

The efforts in understanding a holographic prescription to flatspace have taken two major avenues. The first one, called Celestial holography, proposes that the dual theory to four dimensional (4d) asymptotically flat spacetimes is a 2d relativistic CFT living on the celestial sphere at null infinity, called Celestial CFTs. Following initial observations of Strominger and collaborators \cite{Strominger:2013jfa, He:2014laa, Strominger:2014pwa}, this has led to an impressive body of work relating scattering amplitudes with asymptotic symmetries and memory effects in a web of new relations. The reader is pointed to the excellent reviews \cite{Strominger:2017zoo, Pasterski:2021rjz, Raclariu:2021zjz} for an idea of this rapidly developing field. 

\medskip

The other direction is called Carrollian holography. Here the central idea, perhaps more in keeping with the original holographic principle, is that the field theory dual to gravity in asymptotically flat spacetimes lives one lower dimension and on the entire null boundary $\mathscr{I}^\pm$ of flat space. The point to emphasise is that in the Carrollian proposal, the field theory does not reside just on the celestial sphere, but also on the null direction given by the advanced (or retarded) time coordinate \cite{Bagchi:2016bcd}. The asymptotic symmetries of flat spacetime, called Bondi-van der Burg-Metzner-Sachs (BMS) symmetries \cite{Bondi:1962px, Sachs:1962zza} are encoded on the lower dimensional theory as conformal symmetries, but now on the null surfaces $\mathscr{I}^\pm$. This degenerate metric structure on the boundary necessitates the introduction of new non-metric structures called Carrollian structures, which are mathematically fibre-bundles and the conformal symmetries on null surfaces are Conformal Carrollian symmetries. The putative field theory duals to gravitational theories in 4d asymptotically flat spacetimes are thus 3d Carrollian CFTs. 

\medskip

After the initial suggestion that flat space holography should come from non-Lorentzian symmetries \cite{Bagchi:2010zz,Bagchi:2012cy}, a body of work developed in the case of 3d asymptotically flat spacetimes and 2d field theories where various quantities were computed and shown to match between the bulk and boundary theories. These included the thermal entropy \cite{Bagchi:2012xr, Barnich:2012xq}, correlation functions of stress tensors \cite{Bagchi:2015wna}, entanglement entropy \cite{Bagchi:2014iea, Jiang:2017ecm, Hijano:2017eii}, characters and one-loop determinants \cite{Barnich:2015mui}, asymptotic three point functions \cite{Bagchi:2020rwb}. Other important developments included \cite{Bagchi:2012yk, Barnich:2012aw, Hartong:2015usd}. 

\medskip

The important question of how to connect conformal Carroll correlation functions with scattering matrix elements in one higher dimension was addressed recently in \cite{Bagchi:2022emh}, inspired by Celestial holography \cite{Pasterski:2017kqt, Pasterski:2016qvg} and building on earlier observations in \cite{Banerjee:2018gce}. Further developments and connections to Celestial holography have been recently reported in \cite{Donnay:2022aba, Donnay:2022wvx}. 

\subsection*{From AdS to Flatspace}
One of the ways to arrive at flat spacetimes is to take the infinite radius limit of AdS spacetimes. In the same vein, if one attempts to take a large radius limit of AdS/CFT, one should arrive at a holographic correspondence for flat spacetimes. At the level of symmetries, it was shown that this limit of sending the AdS radius to infinity sends the speed of light on the boundary CFT to zero \cite{Bagchi:2012cy}. Conformal Carrollian structures thus arise naturally. This was one of the principal reasons for the initial formulation of what we now call Carrollian holography. 

\medskip

There has been a lot of work in attempting to understand scattering in asymptotically flat spacetime from taking suitable limits of Witten diagrams in AdS. (We review this in Sec.~3.) It is thus very natural to wonder how one can understand arriving at conformal Carrollian correlation functions as a limit of AdS Witten diagrams and this is the question we answer in this paper. 

\medskip

A particularly intriguing recent piece of work was the elucidation of the connection between AdS$_4$ Witten diagrams in and 2d Celestial correlation functions \cite{PipolodeGioia:2022exe}. We draw inspiration from this work in our paper and show that with a suitable generalisation of this construction, where we explicitly keep track of the null retarded time coordinate while taking the singular limit, we recover 3d Carrollian CFT correlation functions. This provides some tantalising hints at a direct connection between 2d Celestial CFTs and 3d Carrollian CFTs. It is possible that a 2d Celestial CFT is a 3d Carroll CFT restricted on a particular point on the null retarded time direction or is a suitable dimensional reduction of the higher dimensional Carrollian theory.   

\medskip

It is natural that the limit of AdS$_4$ Witten diagrams yields 3d Conformal Carrollian correlators. When one takes a limit of AdS$_4$/CFT$_3$, one does not expect that the boundary theory would suddenly go down in dimension. Since the radius of AdS is taken out to infinity, it is expected that something singular would happen to the boundary theory. The singular limit manifests itself as the vanishing of the speed of light. It has been shown that BMS symmetries of asymptotically flat spacetimes are isomorphic to conformal Carrollian symmetries \cite{Duval:2014uva}. So the matching of (infinite dimensional) symmetries of the boundary theory and the theory in the bulk is manifest in this formulation of Carrollian holography.  

\medskip 

One can think of reaching the null boundary of asymptotically flat spacetimes by performing an infinite boost on the timelike boundary of AdS. The procedure of boosting a CFT on a timelike slice to get a Carrollian CFT on a null manifold in the limit of infinite boosts has been explicitly demonstrated recently for the case of $d=2$ \cite{Bagchi:2022nvj}. Generalisations to higher dimensions should work in analogous ways. 

\medskip  

It has been recently shown that Carrollian hydrodynamics captures the physics of fluids flowing at the speed of light \cite{Bagchi:2023ysc}. This is of course a very high energy limit of usual hydrodynamics. Flat spacetimes have also been thought of as a very high energy limit of AdS, where the Minkowski diamond sits at the centre of a very large radius AdS spacetime and the observers in this region do not feel the curvature of AdS. The fact that Carrollian symmetries can arise as the very high energy subsector of relativistic symmetries, as is manifest from the fluid example, ties in well with this intuition. 

\subsection*{A brief outline of this paper}
As advertised above, the main goal of this paper is to systematically derive Carrollian CFT correlation functions from AdS Witten diagrams. We will do this by generalising the construction of \cite{PipolodeGioia:2022exe} by carefully keeping track of the null direction that was previously not taken into account. For the purposes of this paper, we will focus on Witten diagrams in AdS$_4$ and Carrollian CFT in $d=3$. 

\medskip

To begin, in Sec.~2, we briefly review Carrollian and conformal Carrollian symmetries in arbitrary dimensions before focussing on $d=3$. We revisit Carrollian Ward identities and remind the reader that there are two distinct branches for the correlation functions which are the delta-function branch and the CFT branch. The delta-function branch is the one which is of interest to us in this work as it is related to flatspace scattering \cite{Bagchi:2022emh} via the modified Mellin transformations that were first introduced in \cite{Banerjee:2018gce}. We review this briefly. We then explicitly construct non-trivial three point functions in the collinear limit. This is new material. Appendix \ref{ap:modifiedmellin}, which contains a review of modified Mellin transformations and explains how this is related to Carrollian correlators, supplements this section.  

\medskip

The main message of the paper is contained in Sec.~3, where we show how to go from AdS Witten diagrams to $n$-point functions of Carroll CFTs by keeping track of the null direction in the limit. Thereafter, in Sec.~4, we consider the two and three point Witten diagrams and show explicitly how these reproduce the Carrollian answers derived earlier from Ward identities. We find that the generic three-point function vanishes as expected, but we recover the non-trivial answer arrived at from symmetries in the collinear limit. We end with discussions in Sec.~5. Three appendices give details of computations that are omitted or shortened in the main text.

\newpage

\section{Carrollian Symmetries, Ward Identities and Correlators}
In this section we start off with a quick review of Carrollian and conformal Carrollian symmetries and then focus on the case of $d=3$, where the BMS$_4$ or equivalently 3d Conformal Carroll symmetries arise. We then consider Carrollian Ward identities and review the construction of two point correlation functions, where two distinct branches appear. Three point functions generically vanish, but we find that there are non-trivial conformal Carroll three point functions when we take momenta to be collinear. 

\subsection{Carrollian and Conformal Carrollian Symmetries}

\subsection*{\em{Algebraic point of view}}
The Carroll algebra appears in the vanishing speed of light ($c \to 0$) limit of the Poincare algebra \cite{LevyLeblond, NDS}. In a similar vein, the In{\"o}n{\"u}-Wigner contraction of the relativistic conformal algebra gives the Conformal Carroll algebra. When viewed on coordinates, this is a contraction of the time direction \cite{Bagchi:2012cy} 
\be{}
x^i \to x^i, \quad  t \to \e t, \quad \e \to 0
\ee
where $\e$ is a dimensionless version of the speed of light. This leads to changes in the algebra, e.g. Lorentz boosts become commuting Carrollian boosts:
\begin{subequations}
\bea{}
&& J_{0i} = x_i \partial_t + t \partial_i \Rightarrow C_i = \lim_{\e\to0} \e J_{0i} = x_i \partial_t \\
&& [J_{0i}, J_{0j}] = - J_{ij} \Rightarrow [C_i, C_j] = 0. 
\eea 
\end{subequations}
The relativistic conformal algebra in $d$ spacetime dimensions is $so(d,2)$. The conformal Carrollian algebra in $d$ spacetime dimensions is a contraction of the time direction of $so(d,2)$ and is thus given by $iso(d,1)$. If we remember that the relativistic conformal algebra is isomorphic to the isometry algebra of AdS$_{d+1}$, i.e. $so(d,2)$, then it is obvious that the conformal Carrollian algebra in $d$ spacetime dimensions is isomorphic to the isometry algebra of $(d+1)$ dimensional Minkowski spacetime, i.e. $iso(d,1)$. This is the first rudimentary check for a holographic duality. The isometry algebra of the higher dimensional bulk theory should match with the lower dimensional boundary theory. The Carrollian version of flat holography is based on this simple but fundamental fact. 

\subsection*{\em{Geometric point of view}}
Carrollian symmetries in an arbitrary dimension $d$ arise as the isometries of $d$ dimensional Carroll manifolds, which are non-Lorentzian manifolds endowed with a tensor field $h_{\mu\nu}$ of rank $(d-1)$ and signature $(0, +, +, \ldots +)$ and a no-where vanishing vector $\tau^\mu$ that generates the kernel of $h$ \cite{Henneaux:1979vn, Duval:2014uoa, Duval:2014uva}:
\be{}
h_{\mu\nu}\t^\nu = 0.
\ee 
The Lie algebra of vector fields $\chi = \chi^\mu \partial_\mu$ that generates the isometry algebra of this structure: 
\be{}
\pounds_\chi h_{\mu\nu} = 0, \quad \pounds_\chi \t^\nu = 0 \, ,
\ee
generates an infinite dimensional analogue of the Carroll algebra. This can be made finite by demanding a connection on the Carroll manifold that is compatible with the pair $\left(h_{\mu\nu}, \t^\mu \right)$. This algebra matches the one which we discussed earlier as arising out of the contraction of the Poincare algebra. 

\medskip

If we now consider conformal isometries of the Carroll manifolds:
\be{}
\pounds_\chi h_{\mu\nu} = \lambda_1 h_{\mu\nu}, \quad \pounds_\chi \t^\nu = \lambda_2 \t^\nu \, ,
\ee
this algebra again closes to what is the infinite version of the conformal Carroll algebra. For the case where
\be{}
N= -\frac{\lambda_1}{\lambda_2} = 2 \, ,
\ee
the $d$-dimensional conformal Carroll algebra can be shown to be isomorphic to the BMS$_{d+1}$ algebra, which arises as asymptotic symmetries of $(d+1)$ dimensional flat spacetime at its null boundary \cite{Duval:2014uva}. 

\medskip

It is good to pause here to understand the significance on the last statement. The isometry of Minkowski spacetime is of course the Poincare algebra, and we stressed that the finite version of the Conformal Carroll group is isomorphic to the Poincare group, and thus realises the symmetries of Minkowski spacetime holographically in one lower dimension. One of the surprises of the analysis of Bondi, van der Burg, Metzner \cite{Bondi:1962px} and Sachs \cite{Sachs:1962zza}  in the 1960's was the infinite enhancement of symmetries at the null boundary of 4d asymptotically flat spacetimes. The algebra, called the BMS$_4$ algebra after them, reads 
\begin{subequations}\label{bms4}
\bea{}
&& [L_n, L_m] = (n-m) L_{n+m}, \quad [{\bar{L}}_n, {\bar{L}}_m] = (n-m) {\bar{L}}_{n+m} \\
&& [L_n, M_{r,s}] = \left(\frac{n+1}{2} - r\right) M_{n+r,s}, \quad  [{\bar{L}}_n, M_{r,s}] = \left(\frac{n+1}{2} - s\right) M_{r,n+s} \\
&& [M_{p,q}, M_{r,s}] = 0
\eea
\end{subequations}
Here $M_{r,s}$ are angle dependent translations of the retarded time direction called supertranslations and $L_m$ and ${\bar{L}}_n$ are the infinite dimensional conformal generators on the sphere $\mathbb{S}^2$ at infinity. $L_{0, \pm1}$ and ${\bar{L}}_{0, \pm1}$ constitute the Lorentz algebra, and $M_{r,s}$ for $r,s=0,1$ are the four translation generators. The infinite dimensional supertranslations were found in the original BMS work, but the enhancement of the Lorentz generators to the infinite dimensional superrotations are due to the work of Barnich and Troessaert \cite{Barnich:2010eb} more recently. There are other versions of infinite extensions to include the whole Diff($\mathbb{S}^2$) \cite{Campiglia:2014yka}, but we will not be interested in this here. {\footnote{For issues of the enhancement to Diff($\mathbb{S}^2$) with scattering in flatspace, please have a look at \cite{Schwarz:2022dqf}. From a Carrollian perspective, look at \cite{Dutta:2022vkg}.}}

\medskip

A particularly important message here is that the proposed dual 3d Carrollian CFT to asymptotically 4d flat spacetime naturally encodes this infinite dimensional BMS algebra found independently by the canonical analysis in the bulk, as its symmetries, because the manifold on which it is defined realises the algebra as its conformal isometries. Carrollian CFTs are generically CFTs defined on null surfaces and hence natural co-dimension one holographic dual of asymptotically flat spacetimes. 

\medskip

We should mention here that recently the study of tree-level massless scattering in the bulk flat spacetime has led to the discovery of a much larger symmetry than the BMS group \cite{Banerjee:2020zlg,Strominger:2021mtt}. Thus the asymptotic symmetries, coming from the point of view of scattering, is much richer than the ones mentioned above. We don't yet understand how these additional symmetries show up in the Carrollian framework, but there seem to be some tantalising hints, which we will not discuss further in this work.

\subsection{3d Carroll CFTs}
Our focus, as stated above, is on Carroll CFTs in 3d. The algebra of interest for these theories is the conformal Carrollian algebra or equivalently the BMS$_4$ algebra \refb{bms4}. The geometry of $\mathscr{I}^+$ is that of a fibre bundle with a structure $\mathbb{R}_u \times \mathbb{S}^2$, where $\mathbb{R}_u$ is a null line and the $\mathbb{S}^2$ is the celestial sphere at each point $u$ on the null line. The metric is given by 
\be{}
ds^2_{ \mathscr{I}^+} = 0. du^2 + d\Omega_2^2.
\ee
The BMS$_4$ algebra is the conformal isometry of this degenerate metric, as we have discussed above. A useful representation for this algebra is given in terms of stereographic coordinates on the sphere $z, \bar{z}$ and the null retarded time $u$, where the generators take the form:
\be{}
L_n = -z^{n+1}\partial_z - \frac{1}{2}(n+1)z^n u\partial_u, \, \bar{L}_n = -\z^{n+1}\partial_\z - \frac{1}{2}(n+1)\z^n u\partial_u, \, M_{r,s} = z^r \z^s \partial_u.
\ee
The Carrollian CFT defined on this structure has fields $\Phi(u, z, \z)$ that are characterised by their weights under $L_0, \bar{L}_0$: 
\be{}
[L_0, \Phi(0)]= h  \Phi(0), \quad [\bar{L}_0, \Phi(0)]= \h  \Phi(0).
\ee
We define Carrollian primaries as ones that are annihilated by positive moded generators:
\be{}
[L_n, \Phi(0)]= 0, \quad [\bar{L}_n, \Phi(0)]= 0, \quad \forall n>0, \qquad [M_{r,s}, \Phi(0)]= 0 \quad \forall \, r \, \text{or} \, s>0.
\ee
The primary operators transform under the conformal Carroll symmetries as 
\begin{subequations} \label{Primary transformations}
\begin{align}
\delta_{L_n}\Phi(u,z,\bar{z})&=z^{n+1}\partial_z \Phi(u,z,\bar{z})+(n+1)z^n(h+\frac{u}{2}\partial_u)\Phi(u,z,\bar{z}),  \\ 
\delta_{\bar{L}_n}\Phi(u,z,\bar{z})&=\z^{n+1}\partial_\z \Phi(u,z,\bar{z})+(n+1)\z^n(\h+\frac{u}{2}\partial_u)\Phi(u,z,\bar{z}),  \\ 
 \delta_{M_{r,s}}\Phi(u,z,\bar{z})&=z^r\bar{z}^s\partial_u\Phi(u,z,\bar{z}) \, .
\end{align}
\end{subequations}
In what follows, we will be interested in computing vacuum correlation functions based on conformal Carroll symmetries. We will consider two and three point functions and for this the Poincare subgroup (or finite conformal Carroll symmetries which come about as the limit of relativistic conformal symmetry) comprising of the ten generators $L_{0,\pm1}$,$\bar{L}_{0,\pm1}$ and $M_{00},M_{10},M_{01}$ and $M_{11}$ will suffice in fixing the functional form upto constant factors. 

\medskip

Below we first review the two point function and the emergence of two distinct branches and then derive a non-trivial three-point function for the so-called delta branch. 

\subsection{Two Point function}
The two point function was constructed and its relation with bulk scattering amplitudes was established in \cite{Bagchi:2022emh}. The Ward identities associated with the supertranslations admit two classes of solutions, usually designated as the `CFT' branch and the `delta-function' branch{\footnote{The existence of two distinct classes of solutions for Carroll Ward identities was also noticed in \cite{Chen:2021xkw, deBoer:2021jej}.}. It turns out that only the delta-function branch is relevant in the context of massless scattering amplitudes in the bulk. Here we review the computation for 2-point function and using similar technologies we fix the 3-point correlation function for different cases. 

\medskip

The vacuum two-point function of two Carrollian primaries $\Phi(u,z,\bar{z})$ and $\Phi'(u',z',\bar{z}')$ with weights $(h,h')$ and $(\bar{h},\bar{h}')$ is denoted by: 
\begin{equation}
G^{(2)}(u,z,\bar{z},u',z',\bar{z}')=\<0|\Phi(u,z,\bar{z})\Phi'(u',z',\bar{z}')|0\>
\end{equation}
We will impose invariance under the global Poincare group to find the form of this correlation function. With respect to the global spacetime translations $M_{r,s}$ for $r,s=0,1$, this correlation function varies as 
\begin{equation}
\delta_{M_{r,s}}G^{(2)}(u,z,\bar{z},u',z',\bar{z}')=\big( z^r\bar{z}^s\partial_u + z'^r \bar{z}'^s \partial_{u'}\big)G^{(2)}(u,z,\bar{z},u',z',\bar{z}')=0 \, ,
\end{equation}
where $r,s = 0,1$. This equation has two independent solutions that give rise to two different classes of correlators. On one hand, one can simply get rid of the $u$ dependence from the correlation functions. In this case the two-point function of the 3d Carrollian CFT becomes same as 2d relativistic CFT primary two-point correlator using invariances under the global part of superrotations \cite{Bagchi:2016bcd}. Hence it is given by
\begin{equation}
G^{(2)}(u,z,\bar{z},u',z',\bar{z}')=\frac{\delta_{h, h'} \delta_{\h, \h'}}{(z-z')^{2h}(\bar{z}-\bar{z}')^{2\bar{h}}}. 
\end{equation}
On the other hand, it is possible to keep the $u$ dependence at the expense of settling for a contact term in the spatial part. Thus 
\begin{equation} \label{f}
G^{(2)}(u,z,\bar{z},u',z',\bar{z}')=f(u-u')\delta^{2}(z-z',\bar{z}-\bar{z}').
\end{equation}
$f(u-u')$ is not constrained by the supertranslations hence remains arbitrary for now. With respect to the superrotations these correlators vary as
\begin{align}
\delta_{L_n}G^{(2)}&(u,z,\bar{z},u',z',\bar{z}')=\big[ (z^{n+1}\partial_z+z'^{n+1}\partial_{z'})\\ \nonumber 
 &+(n+1)\big((hz^n+h'z'^n)+\frac{1}{2}(uz^n\partial_u+u'z'^n\partial_{u'})\big)\big]G^{(2)}(u,z,\bar{z},u',z',\bar{z}').
\end{align}
Similarly for $\bar{L}_n$s we have 
\begin{align}
\delta_{\bar{L}_n}G^{(2)}&(u,z,\bar{z},u',z',\bar{z}')=\big[ (\bar{z}^{n+1}\partial_{\bar{z}}+\bar{z'}^{n+1}\partial_{\bar{z}'}) \\ \nonumber
&+ (n+1)\big((\bar{h}\bar{z}^n+\bar{h}'\bar{z'}^n)+\frac{1}{2}(u\bar{z}^n\partial_u+u'\bar{z'}^n\partial_{u'})\big)\big]G^{(2)}(u,z,\bar{z},u',z',\bar{z}').
\end{align}
For $n=-1$, these equations impose translational invariance in spatial direction, but the expression in (\ref{f}) is already invariant under spatial translations. So these equations don't add anything new.

\medskip

However for $n=0$, these equations are 
\begin{align}
[(z\partial_z+z'\partial_{z'})+(h+h')+\frac{1}{2}(u\partial_u+u'\partial_{u'})]f(u-u')\delta^{2}(z-z',\bar{z}-\bar{z}')=0,
\end{align}
and
\begin{align}
[(\bar{z}\partial_{\bar{z}}+\bar{z}'\partial_{\bar{z'}})+(\bar{h}+\bar{h'})+\frac{1}{2}(u\partial_u+u'\partial_{u'})]f(u-u')\delta^{2}(z-z',\bar{z}-\bar{z}')=0.
\end{align}
After using properties of delta-functions, these two equations become
\begin{align}
(\Delta+\Delta'-2)f(u-u')+(u-u')\partial_uf(u-u')=0,  \quad (\sigma+\sigma')f(u-u')=0,
\end{align}
where $\Delta=(h+\bar{h})$ , is the scaling dimension and $\sigma=(h-\bar{h})$, is 2d spin. The solution of the above equations is 
\begin{equation}
f(u-u')=\delta_{\sigma+\sigma'}(u-u')^{-(\Delta+\Delta'-2)}
\end{equation}
Hence
\begin{equation} \label{Sym-cor}
G^{(2)}(u,z,\bar{z},u',z',\bar{z}')=\delta_{\sigma+\sigma', 0}\frac{\delta^{(2)}(z-z',\bar{z}-\bar{z}')}{(u-u')^{\Delta+\Delta'-2}}
\end{equation}
Once this correlator has this form (\ref{Sym-cor}), the equations for $n=1$ , which impose the special conformal invariance, are also trivially satisfied.

\medskip

The two point correlation function of the delta branch is related to the propagation of a free massless particle in the bulk asymptotically flat spacetime from a point on $\mathscr{I}^-$ to a corresponding point on $\mathscr{I}^+$ through the modified Mellin transformation \cite{Banerjee:2018gce}. We review aspects of the modified Mellin transformation in Appendix \ref{ap:modifiedmellin}. The two-point function in the field theory is equivalent to the free propagation of a massless particle from $\mathscr{I}^-$ to $\mathscr{I}^+$. The appearance of the spatial delta-function in the field theory reflects the fact that in the bulk the particle propagates without changing direction on the celestial sphere. 

\subsection{Three Point functions}

We now consider the three-point function of three Carroll primaries $\Phi_i$, with weights $h_i, \h_i$ ($i=1, 2, 3$): 
\begin{equation}
    G^{(3)}(u_i,z_i,\bar{z}_i)=\<0|\Phi_1(u_1,z_1,\bar{z}_1)\Phi_2(u_2,z_2,\bar{z}_2)\Phi_3(u_3,z_3,\bar{z}_3|0\>.
\end{equation}

The Ward identities associated with the three point functions are obviously more involved and allows different sub-classes of solutions. Especially  different solutions in the delta function branch are very subtle and relevant  for different physical scenarios. However for  $u$-independent or the CFT branch the solution again  reduces to the 3-point correlator of a 2d CFT just like the 2-point case \cite{Bagchi:2016bcd}. This is given by
\begin{align}
G^{(3)}(u_i,z_i,\bar{z}_i)=\frac{c_{123}}{z_{12}^{h_1+h_2-h_3}z_{23}^{h_2+h_3-h_1}z_{31}^{h_3+h_1-h_2}}\times \text{complex conjugate}
\end{align}

\medskip

The solutions corresponding to the delta function branch are generically trivial and reflect the fact that in Minkowski space the 3-point scattering amplitudes for massless particles are zero due to momentum conservation \cite{Banerjee:2018gce}, except for in specific cases. The non trivial answers corresponding to either collinear scattering of massless particles, where all 3 particles are parallel to each other or when one of the particles has zero energy. This was pointed out in \cite{Chang:2022seh}, in the context of celestial holography. Using these non-trivial 3-point amplitudes, the authors compute the the 3-point functions of 2d celestial primaries. Here  we study the correlation functions of 3d Carroll CFT primaries, keeping in mind the above mentioned scenarios from symmetry considerations. In this work, we will only consider collinear scatterings and will keep the soft three point functions for future work. 

\medskip

To focus on the collinear case, we begin with the ansatz
\begin{align}
G^{(3)}(u_i,z_i,\bar{z}_i)=F(u_i)\delta^2(z_{12})\delta^2(z_{13}).
\end{align}
The delta function in $z_{12}$ and $z_{13}$ only allows the momenta that are parallel to each other. Global generators of the algebra will further fix the form of $F(u_i)$.

\medskip

Supertranslation invariance would imply
\begin{align}
\sum_{i=1}^3z^{r}_i{z}^s_i\partial_{u_i}G^{(3)}(u_i,z_i,\bar{z}_i)=0  \qquad \forall r,s =0,1.
\end{align}
 Because of the presence of the delta function in the ansatz all these supertarnslation equations effectively become
 \begin{align}
 \sum_{i=0}^3 \partial_{u_i} F(u_i)=0.
 \end{align}
 
This equation implies translation invariance along $u$ direction, i.e. 
\begin{align} \label{series}
F(u_i)\equiv F(u_{12},u_{23},u_{31})=\sum_{a,b,c} f_{abc}u_{12}^a u_{23}^b u_{31}^c.
\end{align}
%

There are still six more equations that would impose global superrotation invariance, which is isomorphic to bulk Lorentz group. Variations with respect to the holomorphic and anti-holomorphic generators are given by 
 \begin{subequations}
 \begin{align}
     \delta_{L_n} G^{(3)}(u_i,z_i,\bar{z}_i)=\sum_{i=1}^{3} \Big[z_{i}^{n+1}\partial_{z_{i}}+(n+1)z_{i}^{n}(h_i+\frac{1}{2}u_i\partial_{u_i})\Big]G^{(3)}(u_i,z_i,\bar{z}_i) \quad \text{and} \\ 
     \delta_{\bar{L}_{n}} G^{(3)}(u_i,z_i,\bar{z}_i)=\sum_{i=1}^{3} \Big [\bar{z_i}^{n+1}\partial_{\bar{z}_{i}}+(n+1)\bar{z_i}^{n}(\bar{h}_{i}+\frac{1}{2}u_{i}\partial_{u_i})\Big]G^{(3)}(u_i,z_i,\bar{z}_i).
     \end{align}
         \end{subequations}
We need to solve these equations for $n=0,\pm 1$. We shall do that for an arbitrary term in the series expansion given by (\ref{series}), i.e.
\begin{equation}
G^{(3)}(u_i,z_i,\bar{z}_i)= u_{12}^a u_{23}^b u_{31}^c \delta^2(z_{12})\delta^2(z_{13}).
\end{equation}
The $n=-1$ equations only imposes translational invariance along $z$ and $\bar{z}$. The ansatz we work with already satisfy these conditions. The equation for $n=0$ reads
\begin{align}
&\sum_{i=1}^{3} \Big[z_{i}\partial_{z_{i}}+z_{i}(h_i+\frac{1}{2}u_i\partial_{u_i})\Big] F(u_i)\delta^2(z_{12})\delta^2(z_{13})=0,  \\ 
&[(-2+\sum_{i=0}^3 h_i)+\frac{1}{2}(a+b+c)]G^3(u_i,z_i,\bar{z}_i)=0.  \\ 
\text{This implies}\quad & a+b+c=-2(-2+\sum_{i}h_i).
\end{align}
Similarly the anti holomorphic equation fixes the anti-holomorphic weights as
\begin{align}
a+b+c=-2(-2+\sum_i\bar{h}_i).
\end{align}
It can be shown that the equations imposed by special conformal generators $L_1$ and $\bar{L}_1$ don't further add any constraints on the solutions. Together the above two equations become
\begin{align}\label{eq:threepointpowers}
a+b+c=4-\sum_{i}\Delta_i, \qquad \sum_{i}\sigma_i=0.
\end{align}

Thus finally we have 
\begin{align}\label{eq:threepointintrinsic}
G^{(3)}(u_i,z_i,\bar{z}_i)=\sum_{b,c} f_{bc}u_{12}^{4-b-c-\sum_{i}\Delta_i}u_{23}^bu_{31}^c \delta^2(z_{12})\delta^2(z_{13}).
\end{align}
Later in the paper we shall show this correlation function captures the collinear scatterings in Minkowski space.

\medskip

Finally, let us also mention that if we assume $z$ and $\bar{z}$ to be independent, then the three point function can have a different non-trivial subclass of solutions. This amounts to working on a celestial torus instead of a celestial sphere. This situation is relevant for (2,2) signature bulk spacetimes \cite{Atanasov:2021oyu}. For these classes of spacetimes, the momentum of the particles are allowed to be complex. Thus there are non-trivial 3-point scattering amplitudes. We will not discuss these cases in this work.

\bigskip

\newpage

\section{Flat Limit of AdS Witten Diagrams}
\label{sec:wittendiagram}

In this section, we elucidate the procedure of taking the limit on AdS Witten diagrams \cite{Witten:1998qj} so as to be able to reach the field theory living on the null boundary of the resulting Minkowski spacetime. The main emphasis again is that the null direction of the boundary Carrollian field theory should play a pivotal role in the whole programme and taking a large radius limit of AdS$_4$ should not lead to a boundary theory which is a codimension two theory, but rather codimension one.


\subsection{A brief history}



Before we go on to describing our methods, it is good to take a look back at some important previous work in this direction. The idea of obtaining S-Matrices from AdS/CFT dates back to the original work of Polchinski \cite{Polchinski:1999ry} and Susskind \cite{Susskind:1998vk}. They first sketched a prescription to obtain flat space S Matrices in the large AdS radius limit. The work \cite{Giddings:1999jq} elaborated on this prescription to relate flat space S-Matrices to analogous construction of scattering amplitudes in AdS \cite{Balasubramanian:1999ri,Giddings:1999qu} \footnote{The notion of scattering amplitudes in AdS is subtle because it is tricky to define wavepackets on the boundary of AdS \cite{PhysRevD.94.065017}. Roughly this is because AdS has a timilike boundary (unlike the Minkowski case) and one cannot define \textit{in} and \textit{out} states.}. Most importantly this prescription gave a way to obtain Minkowski space in the center of AdS by a careful rescaling of the bulk coordinates of AdS. The subtleties in the prescription were refined in \cite{Gary:2009ae} where they show that one can obtain the bulk S-Matrices only if the boundary correlators have a particular singularity 
structure. 

\medskip

These observations were generalized in \cite{Penedones:2010ue} where it was noticed that the Mellin space representation \cite{Mack:2009mi,Mack:2009gy} of the correlation functions of the boundary CFT can be re-interpreted as AdS scattering amplitudes. For massless particles, it was conjectured \cite{Penedones:2010ue} that the bulk S-Matrix is encoded in the large AdS radius limit of the AdS scattering amplitude. This was explicitly checked by computing the Witten diagrams of general scalar theories at tree level and $\phi^4$ theory at one loop level. This conjecture was explicitly proved in \cite{Fitzpatrick:2011jn,Fitzpatrick:2011hu,Fitzpatrick:2011dm} by carefully constructing scattering states in AdS that reduced to the usual plane wave basis in the center of AdS. These states correspond to special primary states in the dual CFT description, thereby relating the CFT correlation functions to flat space scattering in the center of AdS. A generalisation for massive scattering particles was addressed in \cite{Paulos:2016fap}. A flat space limit of Witten diagrams in the momentum space was presented in \cite{Raju:2012zr}.

\medskip

Another parallel approach to obtain bulk S Matrices through the flat space limit of   AdS-CFT is by using HKLL Bulk reconstruction techniques \cite{Hamilton:2005ju,Hamilton:2006az,Hamilton:2006fh}. This approach has been pursued in  \cite{Hijano:2019qmi,Hijano:2020szl,Li:2021snj}. In this approach, one first identifies the Minkowski space in the center of AdS in large AdS radius limit following \cite{Giddings:1999jq}. Within this ``scattering region", one reconstructs the bulk fields in the form of smearing of the CFT operators following HKLL prescription. Since the bulk fields have a mode expansion in terms of the creation and annihilation operators, one obtains scattering states by taking a Fourier transform of the HKLL smearing kernel in the large AdS radius limit. These states are a generalization of the scattering states constructed in \cite{Fitzpatrick:2011jn}. Thus, the analysis of \cite{Hijano:2019qmi} naturally reduces to that of \cite{Penedones:2010ue} when the scattering particles were massless, and it also encompasses \cite{Paulos:2016fap} when the scattering particles were massive. \cite{Li:2021snj} also noted how the flat space limit of momentum space correlators in \cite{Raju:2012zr} can be understood from the point of HKLL Bulk reconstruction techniques.

\medskip

In this formulation, one uses the LSZ formula for S-Matrix within the ``scattering region" and then reconstruct the off-shell bulk correlation functions in terms of the CFT correlators semared over the boundary \cite{Li:2021snj}. The crucial input here is that the smearing regions of the CFT operators were localized around certain time slices of the boundary in the large AdS radius limit. Depending on the choice of the regions, we either get massless scattering or massive scattering. In the global AdS, the operators were smeared around particular future and past time slices given by 
\be{}
\tau = \pm\frac{\pi}{2}+\mathcal{O}(R^{-1})
\ee 
to get massless scattering states. An interpretation of this would be that the CFT data that is required to compute S-Matrices for massless particles is encoded in the region localized around $\tau = \pm\frac{\pi}{2}$. This point would be crucial for us when we reproduce Carrollian CFT correlation functions from the large AdS limit of AdS Witten diagrams. 

\medskip

Recent attempts to link flat holography to AdS/CFT using Witten diagrams are \cite{Casali:2022fro,PipolodeGioia:2022exe,Iacobacci:2022yjo,Sleight:2023ojm}. The connection between Celestial amplitudes and Witten diagrams was pointed out in \cite{Lam:2017ofc} where the feautures of the bulk point singularity in AdS Witten diagrams \cite{Gary:2009ae,Penedones:2010ue} were seen in Celestial amplitudes of two lower dimensions. Then in particular examples, \cite{Casali:2022fro} were able to obtain Celestial amplitudes in 4D flat space from $AdS_3$ Witten diagrams. 

\medskip

The relation between Celestial amplitudes and AdS Witten diagrams in two higher dimensions was elaborated in \cite{PipolodeGioia:2022exe} and this serves as a major inspiration for our work. In \cite{PipolodeGioia:2022exe}, they gave a prescription to obtain Celestial amplitudes from AdS Witten diagrams with boundary correlators inserted at specific past and future slices \footnote{The intuition comes from the bulk reconstruction perspective of \cite{Hijano:2019qmi}.}. 

\subsection{Our prescription: keeping track of null retarded time}
Our main motivation behind reinvestigating the rather well developed literature, as stated in the introduction and above, is that we are not satisfied with the boundary CFT interpretation of the S-matrices in flat spacetimes. On the one hand, we believe that the dual of Minkowski spacetimes should inherit the symmetries of Minkowski spacetime and one should not have to think about this in terms of a CFT dual to some parent AdS which is the point of view of a majority of the literature. On the other hand, we are uncomfortable with the discarding of the null direction in the Celestial CFT picture, since we think that going from AdS to flat space should not result in the dimensional reduction of the dual theory. 

\medskip

It has already been established in \cite{Bagchi:2022emh} that 3d Carrollian CFTs naturally encode S-matrix elements in 4d asymptotically flat spacetimes through the modified Mellin transformation \cite{Banerjee:2018gce}. We will show below that the modified Mellin wave functions \cite{Banerjee:2018gce} which we will call Carrollian wave functions below, naturally arise from the flat limit of AdS Witten diagrams, when the null direction is properly taken into account. 

\medskip

In what follows, we will be inspired by the analysis of \cite{PipolodeGioia:2022exe} which established a general relation between the large AdS radius limit of AdS Witten diagrams and Celestial amplitudes in two lower dimensions. \cite{PipolodeGioia:2022exe} showed that AdS Witten diagrams with boundary CFT operators inserted at specific past and future slices appropriately reduce to the Celestial amplitudes of two lower dimensions in the limit where the AdS radius is large.

\medskip

We however consider a generalisation of the insertion points to obtain higher dimensional structures. Let us point out the major assumptions that we work with when evaluating the Witten diagrams:
\begin{itemize}
	\item The boundary field theory operators are inserted at global time slices $\tau=\pm \frac{\pi}{2}+\frac{u}{R}$.
	\item The two spheres at $\tau= +\frac{\pi}{2} + \frac{u}{R}$ and $\tau= -\frac{\pi}{2} + \frac{u}{R}$ on the boundary are antipodally identified.
\end{itemize}
This should be contrasted with \cite{PipolodeGioia:2022exe} where they work with:
\begin{itemize}
	\item The boundary CFT operators are inserted at global time slices $\tau=\pm \frac{\pi}{2}$.
	\item The two spheres at $\tau= \frac{\pi}{2}$ and $\tau = - \frac{\pi}{2}$ on the boundary are antipodally identified.
\end{itemize}
As we will see, this will significantly change the analysis of \cite{PipolodeGioia:2022exe}.

\medskip

The computation of Witten diagrams is greatly simplified in the embedding space formalism \cite{Dirac:1936fq,Penedones:2007ns}. To begin with, we will start by analysing the individual building blocks of the AdS Witten diagram namely the external lines, vertices and the internal lines in the spirit of \cite{PipolodeGioia:2022exe}. We will carefully implement the large R limit in the individual blocks of the Witten diagrams such that resulting diagram evaluates to the Carrollian CFT correlation functions. Before we begin this analysis we collect the important elements of the embedding space representation which will be used in the next subsequent sections. Further details on the embedding space representation can be found in Appendix \ref{ap:embedding}.

\medskip

The $d$-dimensional AdS solution is described in terms of embedding space coordinates as:
\begin{equation}\label{embedding}
	-(X^0)^2-(X^1)^2 + \sum_{i=2}^{d}(X^i)^2= -R^2 \, ,
\end{equation}
where the embedding space is a $\mathbb{R}^{1,1}\times\mathbb{R}^{1,d-2} $ manifold endowed with the metric:
\begin{equation}\label{eq:embedmetric}
	ds^2 = -dX^+ dX^- -(dX^1)^2+ \sum_{i=2}^{d-1}(dX^i)^2 \, .
\end{equation}
Here $X^{\pm}$ are given by :
\begin{equation}
	X^{\pm}= X^0\pm X^d \, ,
\end{equation}
which are associated with the $\mathbb{R}^{1,1}$ lightcone coordinates. The AdS solution written in parametric form which in the above coordinate system is given by :
\begin{equation}\label{eq:embeddingcoord}
	\begin{split}
		X^+ &= -\frac{R(\cos\tau -\sin\rho \, \Omega_d)}{\cos\rho} \, , \quad X^-=-\frac{R(\cos \tau+\sin\rho \, \Omega_d)}{\cos\rho} \, ,\\
		X^1&=-\frac{R \sin\tau}{\cos\rho} \, , \quad X^i= R \tan\rho \, \Omega_i ~~~~~ i = 2,\dots,d-1 \, ,
	\end{split}
\end{equation}
where $\sum^{d}_{j=2}\Omega^2_j= 1$ and $\Omega_i$ represents the coordinates on $S^{d-2}$. In these embedding space coordinates, the  AdS metric is given by the usual global AdS metric:
\begin{equation}\label{eq:adsmetric}
	ds^2 =\frac{R^2}{\cos^2\rho}\left(-d\tau^2+d\rho^2+\sin^2\rho \, d\Omega_{S^{d-2}}^2\right) \, ,
\end{equation}
where $\tau \in (-\infty , \infty)$ \footnote{The original range of $\tau$ in the coordinates of eq.\eqref{eq:embeddingcoord} is $\tau \in [-\pi,\pi]$ but we unwarp the AdS by fully extending the time like coordinate and go to the so called covering space of AdS.} and $\rho \in [0,\frac{\pi}{2}]$. The limit of this AdS boundary is approached by taking the limit $\rho\rightarrow\frac{\pi}{2}$. Thus, the boundary of eq.\eqref{eq:embeddingcoord} is given by:
\begin{equation}
	\textbf{p}=\lim_{\rho\rightarrow\frac{\pi}{2}}\frac{1}{2}R^{-1}\cos\rho \, \textbf{X} \, ,
\end{equation}
with $\textbf{p}^2=0$. The boundary coordinates are given by
\begin{equation}\label{eq:boundarycoord}
	\begin{split}
		P^+ &= -\dfrac{1}{2}(\cos\,\tau_p - \Omega^d_p) \, , \hspace{2cm} P^- = -\dfrac{1}{2}(\cos\,\tau_p + \Omega^d_p)\, , \\
		P^1 &= -\dfrac{1}{2}\sin\,\tau_p \, , \hspace{3.25cm} P^i = \dfrac{1}{2}\Omega_i \, .
	\end{split}
\end{equation}

One can implement the flat limit on the AdS solution eq.\eqref{eq:adsmetric} by following \cite{Giddings:1999jq}. We set 
\begin{equation}\label{eq:largerlimit}
	\tau = \dfrac{t}{R}  ~~~~~~ \rho = \dfrac{r}{R} \, ,
\end{equation}
and then we let $R \to \infty$. In this limit, the embedding space coordinates of eq.\eqref{eq:embeddingcoord} become:
\begin{equation}\label{eq:coordlarger}
	\begin{split}
		X^+ &= -R\left(1 - \dfrac{r}{R}\Omega_4 \right) \, , \hspace{2cm} X^- =-R \left(1 + \dfrac{r}{R}\Omega_4 \right)\, ,\\
		X^1&=-t \, , \hspace{4.18cm}  X^i = r\, \Omega_i\, .
	\end{split}
\end{equation}
Eq.\eqref{eq:largerlimit} also implies that eq.\eqref{eq:adsmetric} becomes
\begin{equation}\label{mflat}
	 ds^2 \xrightarrow{R \to \infty} -dt^2 + dr^2 + r^2 \, d\Omega^2_{S^{d-2}} \, .
\end{equation}

Before we go into the details of the formulation, we would like to motivate our choice for the operator insertions on the boundary theory, i.e. 
\be{tau}
\tau=\pm \frac{\pi}{2}+\frac{u}{R}
\ee
It is clear that one reaches the boundary when $\rho \to \pi/2$ in the coordinates \eqref{eq:adsmetric}. The boundary metric is flat:
\be{}
ds^2_{\text{CFT}} = -d\t^2 + d\Omega_{d-2}^2
\ee
Now the large radius limit of \eqref{eq:adsmetric} gets us to flat spacetime \refb{mflat}. We wish to reach the null boundary of flat spacetime. To be specific, let us choose $\mathscr{I}^+$. An identical analysis holds for $\mathscr{I}^-$. Consider the retarded time 
\be{}
u = t - r = R(\t-\rho). 
\ee
Now, the AdS boundary was reached with $\rho = \pi/2$ and we label the boundary time as $\t = \t_p$. Putting this back, 
\be{}
u = R \left(\t_p - \frac{\pi}{2}\right) \Rightarrow \t_p = \frac{\pi}{2} + \frac{u}{R}
\ee
As a sanity check, we insert this into the boundary metric:
\be{}
ds^2_{\text{bdy}} = -\frac{1}{R^2} du^2 + d\Omega_{d-2}^2
\ee
We see that in the limit of $R\to \infty$ the metric becomes the null or Carrollian metric at $\mathscr{I}^+$:
\be{}
ds^2_{\text{Carroll}} = 0 . du^2 + d\Omega_{d-2}^2
\ee
For reaching $\mathscr{I}^-$, we would need to choose $\tau= - \frac{\pi}{2}+\frac{v}{R}$ where $v$ is the advanced time $v=t+r$. This justifies our choice \refb{tau}. 

\begin{figure}[t]

\begin{subfigure}{0.5\textwidth}
\includegraphics[width=1.0\linewidth, height=6cm]{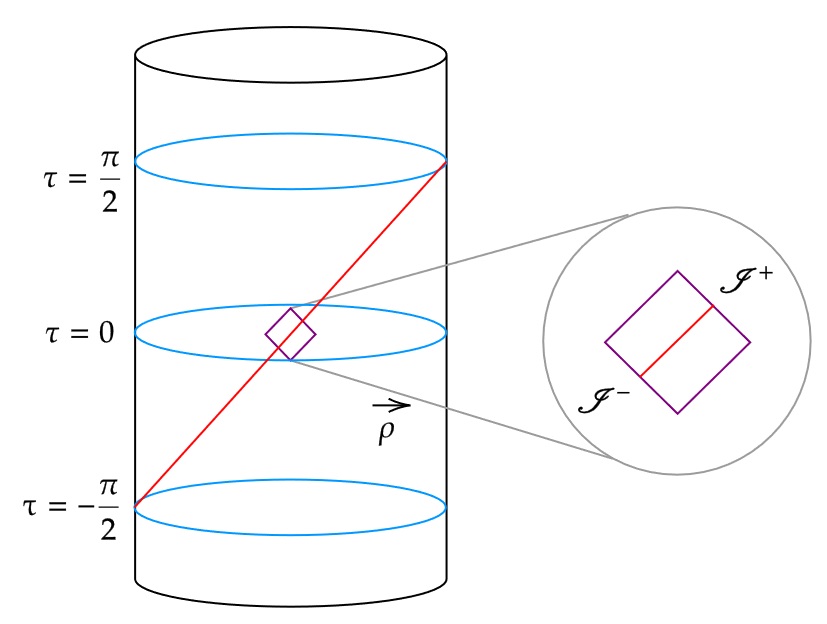} 
\caption{Celestial approach}
\label{fig:cel}
\end{subfigure}
\begin{subfigure}{0.5\textwidth}
\includegraphics[width=1.0\linewidth, height=6cm]{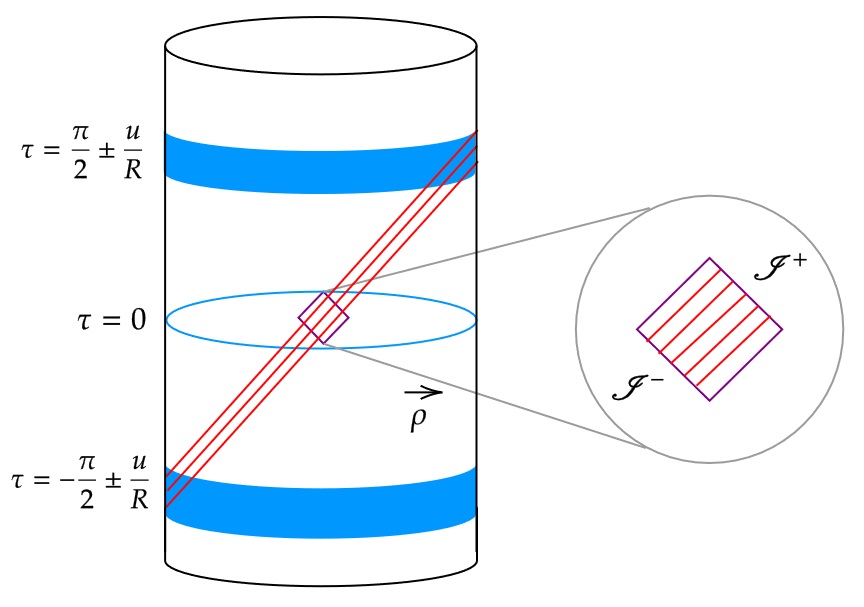}
\caption{Carrollian approach}
\label{fig:car}
\end{subfigure}

\caption{Different approaches to flat limit of AdS}
\label{fig1}
\end{figure}

\medskip

The main differences between our prescription \refb{tau} and that of \cite{PipolodeGioia:2022exe} can be made explicit with Figure \ref{fig1} above. The figures \ref{fig:cel} and \ref{fig:car} represent AdS in global coordinates with the flat Minkowski space (denoted by a purple diamond) arising in the center of AdS in the large $R$ limit. The two different diagrams denote the differences between the Celestial and Carrollian approaches. In the Celestial case drawn in figure \ref{fig:cel}, one works with the fixed time slice $u=0$ (denoted by blue circles in the AdS cylinder) as mentioned above, but in the Carrollian case drawn in figure \ref{fig:car}, one has the entire null boundary (denoted by blue patches in AdS cylinder). The red lines denote the massless particles scattering. The new element in the Carrollian case is that, one gets from the boundary of AdS to the boundary of flat space through \refb{tau} in the large $R$ limit. The large $R$ limit of \refb{tau} is realized as an infinite boost, taking the timelike boundary to a null boundary. The blue patches in the Carroll case of course become vanishingly small and tend to the line in the Celestial diagram as we go to $R\to\infty$, but the smearing of the operators in the next-to-leading $\mathcal{O}(\frac{1}{R})$ becomes all important in keeping the whole null boundary of the asymptotically flat spacetime at the centre of AdS. As we have repeatedly stressed, unlike the Celestial approach, this Carrollian perspective is more natural from the context of holography in that the boundary remains codimension one in this limit.

\subsection{Bulk to boundary propagator}
With the setup of the embedding space formulation, we now begin the analysis of the building blocks of Witten diagrams under a large AdS radius $R$ expansion. We first address the bulk to boundary propagator. 

\medskip

The bulk to boundary propagator $\textbf{K}_\Delta(\textbf{p},\textbf{x})$ in the embedding space is given by \cite{Penedones:2010ue,Penedones:2007ns}: 
\begin{equation}\label{eq:bulkbdyprop}
	\textbf{K}_\Delta(\textbf{p},\textbf{x})=\frac{C_\Delta^d}{(-2\textbf{p}\cdot \textbf{x}+i\epsilon)^\Delta} \, ,
\end{equation}
where
\begin{equation}\label{eq:cddelta}
	C_\Delta^d=\frac{\Gamma(\Delta)}{2 \pi^{\frac{d}{2}}\Gamma(\Delta-\frac{d}{2}+1)R^{\frac{(d-1)}{2}-\Delta}} \, .
\end{equation}
A derivation of this is given in Appendix \ref{ap:embedding}. Here we have parametrized the bulk and boundary points by $\textbf{x}\in(\tau,\rho,\Omega)$ and $\mathbf{p}\in(\tau_p,\Omega_\p)$ respectively.  

\medskip

We now substitute eq.\eqref{eq:largerlimit} in eq.\eqref{eq:bulkbdyprop} and do a series expansion of the bulk to boundary propagator under a large $R$ limit to arrive at
\begin{equation}\label{eq:bbdplargerexp}
	\textbf{K}_\Delta(\textbf{p},\textbf{x})= C_\Delta^d \frac{1}{(R \cos\tau_p+ t \sin\tau_p-r \Omega_\p \cdot \Omega+O(R^{-1})+i\epsilon)^\Delta} \, .
\end{equation}
In arriving at this result, we used eq.\eqref{eq:coordlarger} and eq.\eqref{eq:boundarycoord}. A derivation of eq.\eqref{eq:bbdplargerexp} is given in Appendix \ref{ap:bbdplarger}. Naively, this expression vanishes as $R \to \infty$. To obtain  non-zero leading order contributions to the bulk to boundary propagator we crucially impose the condition
\begin{equation}\label{eq:keycondition}
	\tau_p= \pm\frac{\pi}{2}+\frac{u}{R} \, ,
\end{equation}
where $u$ is a new parameter. This should be contrasted with \cite{PipolodeGioia:2022exe} where they strictly set $\tau = \pm \frac{\pi}{2}$. This constitutes the main deviation of our approach when compared to \cite{PipolodeGioia:2022exe}. Under this assumption, choosing $\tau_p = \frac{\pi}{2}+ \frac{u}{R}$, the bulk to boundary propagator of eq.\eqref{eq:bulkbdyprop} is given by:
\begin{equation}
	\textbf{K}_\Delta(\textbf{p},\textbf{x})= C^d_\Delta \left(\frac{1}{(-u+ t-r\Omega_p \cdot \Omega+i\epsilon)^\Delta}+ O(R^{-1})\right) \, .
\end{equation}
This can be rewritten as 
\begin{equation}\label{eq:outgoingfn}
	\textbf{K}_\Delta(\textbf{p},\textbf{x})= C^d_\Delta \left(\frac{1}{(-u-\tilde{q} \cdot x+i\epsilon)^\Delta}+ O(R^{-1})\right) \, ,
\end{equation}
where $x=(t,r\Omega)\in\mathbb{R}^{1,d}$ is a point in the bulk region and the vector $\tilde{q}=(1,\Omega_p)\in\mathbb{R}^{1,d}$  is a null vector in the direction of the boundary point $\p \in (\tau_p,\Omega_p)$.

\medskip

Similarly for $\tau_p= -\frac{\pi}{2}+\frac{u}{R}$, the bulk to boundary propagator eq.\eqref{eq:bulkbdyprop} under the large R series expansion is given by:
\begin{equation}\label{eq:incomingfn}
	\textbf{K}_\Delta(\textbf{p},\textbf{x})= C^d_\Delta \left(\frac{1}{(u+\tilde{q} \cdot x+i\epsilon)^\Delta}+ O(R^{-1})\right) \, ,
\end{equation}
where $\mathbf{x}$ is the same as the one defined above but the $\tilde{q}=(1,\Omega_\p^A)$, where crucially, we have
\begin{equation}\label{eq:antipodal}
	\Omega_\p^A= -\Omega_\p \, ,
\end{equation}
which is the antipodal point of $\Omega_\p$. The spheres at $\tau_p = \frac{\pi}{2}+\frac{u}{R}$ and $\tau_p =- \frac{\pi}{2}+\frac{u}{R}$ are antipodally matched. This is a generalization of the antipodal matching condition proposed in \cite{PipolodeGioia:2022exe}. Of course, when we set $u=0$, the insertion eq.\eqref{eq:keycondition} and the matching condition eq.\eqref{eq:antipodal} reduces to the one in \cite{PipolodeGioia:2022exe}.

\bigskip

The leading order bulk-to-boundary propagators are in a Carrollian primary basis associated with a modified Mellin transform \cite{Banerjee:2018gce} given by:
\begin{equation}\label{eq:modmellin}
	\int^{\infty}_{0} d\omega \, \omega^{\Delta -1} \, e^{\mp i\omega u} e^{\mp i\omega \, \tilde{q} \cdot x} e^{-\epsilon \omega} = \dfrac{(\pm i)^{\Delta}\Gamma(\Delta)}{(\mp u\mp \tilde{q} \cdot x+i\epsilon)^{\Delta}} \, .
\end{equation}
This is clearly the wavefunctions that we obtain in eq.\eqref{eq:outgoingfn} and eq.\eqref{eq:incomingfn} from the large $R$ limit.

Thus, one can rewrite the bulk to boundary propagators in terms of modified Mellin transforms:
\begin{equation}
	 \textbf{K}_{\Delta}(\p,\x) = N_{\Delta}^{d}\psi_{\Delta,\tilde{q},u}^{\pm}(x) +O(R^{-1}) \, ,
\end{equation}
for
\begin{equation}
	\psi_{\Delta,\tilde{q},u}^{\pm}(x)=\int_{0}^{\infty}d\omega \, \omega^{\Delta_1-1}e^{\mp i \omega (\tilde{q}\cdot x+u)}e^{-\epsilon\omega} \, .
\end{equation}
The normalisation constant $N_{\Delta}^{d}$ is given by
\begin{equation}
	N_{\Delta}^{d} = \frac{(\mp i)^{\Delta}}{2\pi^{\frac{d}{2}}\Gamma(\Delta-\frac{d}{2}+1)R^{\frac{(d-1)}{2}-\Delta}} \, .
\end{equation}
The superscript $+(-)$ indicates outgoing (incoming) wave basis respectively. 

\medskip

To summarize this section, we point out that the key differences when compared to \cite{PipolodeGioia:2022exe} is eq.\eqref{eq:keycondition} and the generalized antipodal matching condition eq.\eqref{eq:antipodal}. This leads us to Carrollian primary wave functions in eq.\eqref{eq:incomingfn} and eq.\eqref{eq:outgoingfn} with explicit $u$ dependence \cite{Banerjee:2018gce} rather than the Celestial Conformal primary wave functions worked out in \cite{Pasterski:2016qvg,Pasterski:2017kqt}. 

\subsection{Vertices}

For the case of non-derivative coupling that we are interested in, the AdS vertices are given by:
\begin{equation}
	i \mu \int_{\text{AdS}_{d+1}} d^{d+1}\x.
\end{equation}

The above measure can be written in the global coordinates and using eq.\eqref{eq:largerlimit} in the large $R$ expansion, we see that the leading order term in the measure is given by:
\begin{equation}\label{eq:measure}
	d^{d+1}\x = d^{d+1}x + O(R^{-2}) \, .
\end{equation}
This is because in the large $R$ expansion, eq.\eqref{eq:adsmetric} implies
\begin{equation}\label{eq:detrel}
	\sqrt{-g_{\text{AdS}_{d+1}}} = \dfrac{R^2}{\cos^2\rho}(R \, \tan\rho)^{d-1} \sqrt{g_{\Omega}} \rightarrow R^2 \, r^{d-1} \, \sqrt{g_{\Omega}} = R^2 \,\sqrt{-g_{\mathbb{R}^{1,d}}} \, .
\end{equation}
Here $g$ denotes the determinant and $g_{\Omega}$ is the determinant of the sphere metric $S^{d-1}$. Also, it is evident from the above expressions that under the large R limit $t(= R \tau)\in(-\infty,\infty)$ and $r(=R\rho)\in\left[0,\infty\right)$.
Hence, we can now define a map from the AdS measure to vertex rule in $ \mathbb{R}^{1,d} $
\begin{equation}
	i \mu \int_{\text{AdS}_{d+1}} d^{d+1}\x = i \mu \int_{\mathbb{R}^{1,d}}\left(d^{d+1}x + O(R^{-2})\right) \, .
\end{equation}

\subsection{Internal lines}

The bulk to bulk propagator in AdS that deals with the operator exchange of dimension $\Delta$ for $\text{AdS}_{d+1}$ satisfies the following equation
\begin{equation}
	\left(\Box_{\text{AdS}_{d+1}}-\frac{\Delta(\Delta-d)}{R^2}\right)\Pi_\Delta(\x_1,\x_2)= i\delta_{\text{AdS}_{d+1}}(\x_1,\x_2) \, .
\end{equation}
The Laplacian when expanded in terms of global coordinates eq.\eqref{eq:adsmetric}, is given by:
\begin{equation}
	\begin{split}
		\Box_{\text{AdS}_{d+1}}&= \frac{-\cos^2\rho}{R^2}\partial_\tau^2+\frac{\cos^{d+1}\rho}{\sin^{d-1}\rho}\partial_\rho\left(\frac{\cos^{d+1}\rho}{\sin^{d-1}\rho}\sqrt{\gamma}\frac{\cos^2\rho}{R^2}\partial_\rho\right)+ \frac{\cos^2\rho}{R^2 \sin^2\rho}\frac{1}{\sqrt{\gamma}}\partial_A\left(\sqrt{\gamma}\gamma^{AB}\partial_B\right)\\
		&=\Box_{\mathbb{R}^{1,d}}+O(R^{-2}) \, ,
	\end{split}
\end{equation}
where $\gamma$ is the metric on $S^{d-1}$ and $\Box_{\mathbb{R}^{1,d}}$ is the flat space Laplacian. Similarly, the delta function can also be written as:
\begin{equation}
	\begin{split}
		\delta_{\text{AdS}_{d+1}}(\x_1,\x_2)&= \frac{\delta(\tau_1-\tau_2)\delta(\rho_1-\rho_2)\delta^{d-1}(\Omega_1-\Omega_2)}{\sqrt{-g_{\text{AdS}_{d+1}}}}\\
		&= \delta_{\mathbb{R}^{1,d}}(x_1,x_2)+O(R^{-2}) \, ,
	\end{split}
\end{equation}
were in the final step, we used eq.\eqref{eq:detrel} and $\delta(a \, x) = \delta(x)/a$. Hence, under the large R expansion of the bulk to bulk propagator is:
\begin{equation}
	\left(\Box_{\mathbb{R}^{1,d}}+O(R^{-2})-\frac{\Delta(\Delta-d)}{R^2}\right)\Pi_\Delta(\x_1,\x_2)= i\delta_{\mathbb{R}^{1,d}}(x_1,x_2)+O(R^{-2}) \, .
\end{equation}
This implies a series expansion of the propagator 
\begin{equation}\label{eq:propagator}
	\Pi_\Delta(\x_1,\x_2)= G(x_1,x_2)+O(R^{-2}) \, ,
\end{equation}
where $G(x_1,x_2)$ obeys the equation 
\begin{equation}\label{eq:flatspacekg}
	\left(\Box_{\mathbb{R}^{1,d}}-m^2\right)G(x_1,x_2)= i\delta_{\mathbb{R}^{1,d}}(x_1,x_2) \, , \quad m\equiv \lim_{R\rightarrow\infty}\frac{\Delta}{R} \, .
\end{equation}
Since $\Pi_\Delta(\x_1,\x_2)$ computes the time-ordered two-point function in AdS, it necessarily means that $G(x_1,x_2)$ has to be the Feynman propagator. Eq.\eqref{eq:flatspacekg} implies that if we are considering massless exchanges in the $R \to \infty$ limit, we should have $\Delta \sim \mathcal{O}(1)$. Presumably, for massive exchanges we should have, $\Delta \sim \mathcal{O}(R)$. We will not have anything to say about massive exchanges in the explicit computations of our paper. 

\subsection{Construction of the diagrams}

Now that we have all the required structures in eq.\eqref{eq:incomingfn}, eq.\eqref{eq:outgoingfn}, eq.\eqref{eq:measure}, eq.\eqref{eq:propagator}, we can combine them to get the complete diagrams; in doing so, we can always have a series expansion of these diagrams under the large R expansion. The term we are interested in is the leading order term.  

\medskip

The explicit form of the propagators in the large $R$ limit are derived to be:
\begin{equation}\label{eq:allpropagatorslarger}
	\begin{split}
		&\text{incoming:} \quad \textbf{K}_{\Delta_{1}}(\p_1,\x) = N_{\Delta_1}^{d}\psi_{\Delta_1,\tilde{q}_1,u_1}^{-}(x) +O(R^{-1}) \,, \\
		&\text{outgoing:} \quad  \textbf{K}_{\Delta_{2}}(\p_2,\y) = N_{\Delta_2}^{d} \psi_{\Delta_2,\tilde{q}_2,u_2}^{+}(y)+{O{(R^{-1})}} \, ,\\
		 &\text{bulk-bulk:} \quad \Pi_\Delta(\x,\y) = G(x,y)+O(R^{-2}) \, ,
	\end{split}
\end{equation}
where,
\begin{equation}
	\begin{split}
		\psi_{\Delta_1,\tilde{q}_1,u_1}^{-}(x)&=\int_{0}^{\infty}d\omega_1 \, \omega_1^{\Delta_1-1}e^{i \omega_1 (\tilde{q}_1 \cdot x+u_1)}e^{-\epsilon\omega_1 }\, , \\
		\psi_{\Delta_2,\tilde{q}_2,u_2}^{+}(y)&=\int_{0}^{\infty}d\omega_2 \, \omega_2^{\Delta_2-1}e^{-i \omega_2 (\tilde{q}_2 \cdot y+u_2)}e^{-\epsilon\omega_2 } \, , \\
		G(x,y)&= \int \frac{d^{d+1}k}{(2\pi)^{d+1}} \frac{e^{ik \cdot (x-y)}}{k^2+m^2+i\epsilon} \, ,
	\end{split}
\end{equation}
and the normalisation constant $N_{\Delta_i}^{d}$ is given by
\begin{equation}
	N_{\Delta_i}^{d} = \frac{(\mp i)^{\Delta_i}}{2\pi^{\frac{d}{2}}\Gamma(\Delta_i-\frac{d}{2}+1)R^{\frac{(d-1)}{2}-\Delta_i}} \, .
\end{equation}
As a quick sanity check, one can see that the incoming and outgoing wave functions of eq.\eqref{eq:allpropagatorslarger} satisfy the flat space $\mathbb{R}^{1,d}$ wave equation \cite{Banerjee:2018gce} by using the representation of eq.\eqref{eq:modmellin}:
\begin{equation}
	\Box_{\mathbb{R}^{1,d}} \textbf{K}_{\Delta}(\p,\x) = N^d_{\Delta} \Box_{\mathbb{R}^{1,d}} \int^{\infty}_{0} d\omega \, \omega^{\Delta -1} \, e^{-i\omega u} e^{-i\omega \, \Tilde{q} \cdot x} \propto - \omega^2 \Tilde{q}^2 =0 \, ,
\end{equation}
as $\Tilde{q}^2 = 0$.

\medskip

A general Witten diagram with appropriate vertices, internal lines and external lines will be of the form
\begin{equation}\label{eq:genwitten}
	\begin{split}
		\langle O_{\Delta_1}(\p_1) O_{\Delta_2}(\p_2) \dots O_{\Delta_i}(\p_i) O_{\Delta_{i+1}}(\p_{i+1}) O_{\Delta_{i+2}}(\p_{i+2}) \dots O_{\Delta_j}(\p_j) \rangle \\
		= (i\mu)^V \int_{\text{AdS}_{d+1}}d^{d+1}\x_1 \dots d^{d+1}\x_V \Pi_{\Delta}(\x_k,\x_l) \dots \Pi_{\Delta}(\x_k',\x_l') \\
		\times \textbf{K}_{\Delta_1}(\p_1,\x_k) \dots \textbf{K}_{\Delta_i}(\p_i,\x_k')
		\textbf{K}_{\Delta_{i+1}}(\p_{i+1},\x_l) \dots \textbf{K}_{\Delta_j}(\p_j,\x_l') \, ,
	\end{split}
\end{equation}
where $V$ denotes the number of vertices and there are appropriate number of external lines connected to the bulk-to-bulk propagator. For example, a four point function with the $\phi^3$ interaction will be of the form
\begin{equation}\label{eq:fourpointeg}
	\begin{split}
		\langle O_{\Delta_1}(\p_1)O_{\Delta_2}(\p_2)O_{\Delta_3}(\p_3)O_{\Delta_4}(\p_4)\rangle= &(i\mu)^2\int_{AdS_{d+1}}d^{d+1}\x \, d^{d+1}\y \, \Pi_\Delta(\x,\y)\\ &~~\textbf{K}_{\Delta_{1}}(\p_1,\x)\textbf{K}_{\Delta_{2}}(\p_2,\x)\textbf{K}_{\Delta_{3}}(\p_3,\y)\textbf{K}_{\Delta_{4}}(\p_4,\y)	\, .
	\end{split}
\end{equation}
Now suppose, $\p_m$ are inserted at $\tau = -\frac{\pi}{2}+\frac{u_m}{R}$ for $m=1,\dots,i$ and at $\tau = \frac{\pi}{2}+\frac{u_n}{R}$ for $n=i+1,\dots,j$, then using the form of the incoming, outgoing wavefunctions, and the bulk to bulk propagators derived in eq.\eqref{eq:allpropagatorslarger}, we can simplify eq.\eqref{eq:genwitten} in the large $R$ limit as
\begin{equation}\label{eq:genwittenlarger}
	\begin{split}
		&\langle O_{\Delta_1}(\p_1) \dots O_{\Delta_i}(\p_i) O_{\Delta_{i+1}}(\p_{i+1}) \dots O_{\Delta_j}(\p_j) \rangle \\
		 &=(\prod^{j}_{a=1}N^d_{\Delta_a})(i\mu)^V \int_{\mathbb{R}^{1,d}} d^{d+1}x_1 \dots d^{d+1}x_V \,  G(x_k,,x_l) \dots G(x_k',x_l') \,
		\psi^-_{\Delta_1,\Tilde{q}_1,u_1}(x_k)  \dots \\
		& \hspace{4cm}\times \psi^-_{\Delta_i,\Tilde{q}_i,u_i}(x_k')\, \psi^+_{\Delta_{i+1},\Tilde{q}_{i+1},u_{i+1}}(x_l)  \dots \psi^+_{\Delta_j,\Tilde{q}_j,u_j}(x_l') + O(R^{-1}) \, .
	\end{split}
\end{equation}
For example in eq.\eqref{eq:fourpointeg}, if $\p_1,\p_2$ are inserted in $\tau = -\frac{\pi}{2}+\frac{u_1}{R}$ and $\tau = -\frac{\pi}{2}+\frac{u_2}{R}$ respectively, and $\p_3,\p_4$ are inserted in $\tau = \frac{\pi}{2}+\frac{u_3}{R}$ and $\tau = \frac{\pi}{2}+\frac{u_4}{R}$ respectively, we get
\begin{equation}
    \begin{split}
        \begin{split}
		\langle O_{\Delta_1}(\p_1)O_{\Delta_2}(\p_2)O_{\Delta_3}(\p_3)O_{\Delta_4}(\p_4)\rangle= &(i\mu)^2\int_{\mathbb{R}^{1,d}}d^{d+1}x \, d^{d+1}y \, G(x,y)\\ &~~ \psi_{\Delta_1,q_1,u_1}^{-}(x) \psi_{\Delta_2,q_2,u_2}^{-}(x) \psi_{\Delta_3,q_3,u_3}^{+}(y) \psi_{\Delta_4,q_4,u_4}^{+}(y)	\, .
	\end{split}
    \end{split}
\end{equation}
We will explicitly evaluate eq.\eqref{eq:genwittenlarger} for specific cases of the two point and the three point correlation function in the next section.

\newpage

\section{Carroll Correlators from Witten Diagrams}

In this section, we will combine all the ingredients of Section \ref{sec:wittendiagram} to construct AdS Witten diagrams in a large $R$ series expansion. We show that the Carrollian CFT correlation functions naturally arise as the leading order term in eq.\eqref{eq:genwittenlarger}.    

\medskip

To compare with the explicit calculations, we will work in $3+1$ dimensions because $\text{BMS}_4$ was used to constrain the correlation functions of the field theory. But it is important to stress that the procedure outlined in the previous section is valid for any dimensions. Returning to the case that we are interested in at the moment, i.e. $3+1$ dimensions, we will use the following parametrization for the null vector in the direction of $\Omega_p$:
\begin{equation}\label{eq:4dparamq}
	\tilde{q}^{\mu} = \left[1,\dfrac{z+\bar{z}}{1+z\bar{z}},\dfrac{-i(z-\bar{z})}{1+z\bar{z}},\dfrac{1-z\bar{z}}{1+ z\bar{z}} \right] \, .
\end{equation}
The topology of the null boundary is $\mathbb{R}\times \mathbb{S}^2$ and $(z,\bar{z})$ characterize the coordinates on the celestial sphere. The role of dimensions come  in when we try to split the momentum conserving delta functions into a delta function in energies and the delta function in the directions. 
We will now consider various $n$-point Witten diagrams in the large AdS radius limit.

\subsection{Two point function}
\label{sec:twopoint}

\begin{figure}
    \centering
    \includegraphics[scale=0.7]{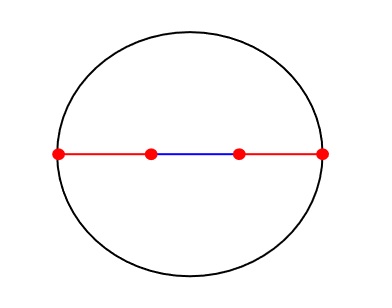}
    \caption{Two point Witten diagram. Red lines denote bulk to boundary propagators; Blue line denotes bulk to bulk propagator.}
    \label{fig:twopoint}
\end{figure}

We begin with the two-point Witten diagram given by
\begin{equation}\label{eq:tpf}
	\langle O_{\Delta_1}(\p_1)O_{\Delta_2}(\p_2)\rangle= \int_{\text{AdS}_{4}}d^{4}\x d^{4}\y\Pi_\Delta(\x,\y) \textbf{K}_{\Delta_{1}}(\p_1,\x)\textbf{K}_{\Delta_{2}}(\p_2,\y)	\, ,
\end{equation}
where $\textbf{K}_{\Delta_{1}}(\p_1,\x)$ is inserted at $\tau= -\frac{\pi}{2}+\frac{u_1}{R}$ (ingoing) and $\textbf{K}_{\Delta_{2}}(\p_2,\y)$ is inserted at $\tau= \frac{\pi}{2}+\frac{u_2}{R}$ (outgoing). This is diagramatically denoted by figure \ref{fig:twopoint} where the red lines denote the bulk to boundary propagators and the blue line denotes the internal bulk to bulk propagator. This is a particular case of eq.\eqref{eq:genwitten}.

\medskip

Thus, the leading order term of eq.\eqref{eq:genwittenlarger} for eq.\eqref{eq:tpf} for the particular insertions is given by
\begin{equation}
	\langle O_{\Delta_1}(\p_1)O_{\Delta_2}(\p_2) \rangle \simeq N^3_{\Delta_1} N^3_{\Delta_2} \int_{\mathbb{R}^{1,3}} d^{4}x \, d^{4}y \, G(x,y) \psi^-_{\Delta_1,\Tilde{q}_1,u_1}(x) \psi^+_{\Delta_2,\Tilde{q}_2,u_2}(y) \, .
\end{equation}
This evaluates to
\begin{equation}\label{eq:twopointfirst}
	\begin{split}
		\langle O_{\Delta_1}(\p_1)O_{\Delta_2}(\p_2)\rangle \simeq&  N_{\Delta_1}^{3}N_{\Delta_2}^{3}\int_{\mathbb{R}^{1,3}}d^{4}x \, d^{4}y \, d\omega_1 \, d\omega_2 \, d^{4}k \, \omega_1^{\Delta_1-1}\omega_2^{\Delta_2-1}e^{-\epsilon(\omega_1+\omega_2)}\\ &\times e^{i\omega_1. u_1} e^{-i\omega_2. u_2} \frac{e^{i(\omega_1\tilde{q}_1+k)\cdot x} e^{-i(\omega_2\tilde{q}_2+k)\cdot y}}{k^2+m^2+i\epsilon} \, .
	\end{split}
\end{equation}
Using the properties of Dirac delta function, and a series of integrations the above expression can be simplified resulting in the final expression given by central result of this paper:
\begin{equation}\label{result1}
	\langle O_{\Delta_1}(\p_1)O_{\Delta_2}(\p_2)\rangle = \mathcal{A} \frac{\delta^2(z_2-z_1)}{(i(u_2-u_1))^{\Delta_1+\Delta_2-2}} \, ,
\end{equation}
where $\mathcal{A}$ is a normalization constant given by
\begin{align}\label{eq:normconst}
	\mathcal{A}=\frac{(2\pi)^{4}N_{\Delta_1}^{3}N_{\Delta_2}^{3}\Gamma(\Delta_1+\Delta_2-2)}{m^2} 
	=\frac{4\pi(-i)^{\Delta_1} i^{\Delta_2}}{m^2} \frac{\Gamma(\Delta_1+\Delta_2-2)}{\Gamma(\Delta_1-\frac{1}{2})\Gamma(\Delta_2-\frac{1}{2})} \frac{1}{R^{2-(\Delta_1+\Delta_2)}} \, .
\end{align}

The detailed calculation is given in Appendix \ref{ap:twopointcalc}. The result \eqref{result1} matches with the  Carrollian CFT two-point correlation function discussed in Sec.~2 (eq.\eqref{Sym-cor}) up to normalisation. 

\medskip

Before we go on to the three point function, let us briefly comment on the normalisation. It seems that in order for the normalisation to be non-zero in the $R\to\infty$ limit, we require $\Delta_1+\Delta_2 =2$. This is reminiscent of the restriction of $\Delta=1$ encountered in \cite{Donnay:2022aba, Donnay:2022wvx}. But we should point out that there is also the mass term $m$ in the bulk-to-bulk propagator which shows up in the normalisation constant and needs to go to zero at the end of the calculation.  Saying something concrete about this constant is thus subtle and requires a more thorough investigation along the lines of adapting holographic renormalisation techniques to flat spacetimes which we will not address in this work. It is very likely that the restriction on the weights would disappear once this is done carefully. Similar consideration would hold for the normalisation of the three-point function which we go on to describe next.   

\subsection{Three point function}
\label{sec:threepoint}

\begin{figure}
    \centering
    \includegraphics[scale=0.7]{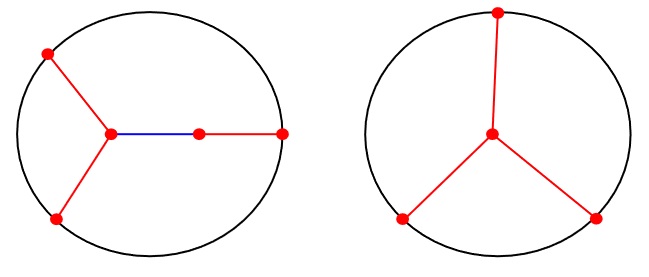}
    \caption{Three point Witten diagrams. Red lines denote bulk to boundary propagators; Blue line denotes bulk to bulk propagator.}
    \label{fig:threepoint}
\end{figure}

To obtain a non-zero three point function, we need a $\phi^3$ interaction in the bulk. There are three possible diagrams that can arise: two incoming particles can exchange an operator with one outgoing particle; one incoming particle can exchange an operator with two outgoing particles; one contact diagram. Interestingly, we find that for generic momentum, the three point function vanishes, which is of course in keeping with the flat space result from momentum conservation. But the nice thing is that we can take a limit of non-zero AdS Witten diagram answers and reach this outcome. In the previous section, we had seen that from symmetry analysis, one can get non-trivial answers for the three point functions when the momenta are collinear. Here we reproduce this answer from a bulk computation starting with an  AdS Witten diagram and taking the suitable limit. 

\subsubsection{Generic momenta}

Let us consider the diagram with two incoming particles exchanging an operator with one outgoing particle. This is diagramatically denoted in figure \ref{fig:threepoint}. The AdS Witten diagram is given by
\begin{equation}\label{eq:threepointflat}
	\begin{split}
		\langle O_{\Delta_1}(\p_1)O_{\Delta_2}(\p_2)O_{\Delta_3}(\p_3) \rangle= &(i\mu)\int_{\text{AdS}_{4}}d^{4}\x \, d^{4}\y\Pi_\Delta(\x,\y)\\ &~~\times \textbf{K}_{\Delta_{1}}(\p_1,\x)\textbf{K}_{\Delta_{3}}(\p_3,\y)\textbf{K}_{\Delta_{2}}(\p_2,\x) \, ,
	\end{split}
\end{equation}
where we will insert $\p_1$ at $\tau=-\frac{\pi}{2}+ \frac{u_1}{R}$, $\p_2$ at $\tau=-\frac{\pi}{2}+ \frac{u_2}{R}$ and $\p_3$ at $\tau=\frac{\pi}{2}+\frac{u_3}{R}$. According to the general analysis of section \ref{sec:wittendiagram} ((i.e.) eq.\eqref{eq:genwittenlarger}), as $R \to \infty$, we have
\begin{equation}\label{eq:threepointfirst}
	\begin{split}
		\langle O_{\Delta_1}(\p_1)O_{\Delta_2}(\p_2)O_{\Delta_3}(\p_3) \rangle \simeq ~ &(i\mu) N^d_{\Delta_1}N^d_{\Delta_2}N^d_{\Delta_3}\int_{\mathbb{R}^{1,3}} \, d^{4}x \, d^{4}y \, G(x,y)\\ &\hspace{1cm}\times \psi^-_{\Delta_1,\Tilde{q}_1,u_1}(x) \,\psi^-_{\Delta_2,\Tilde{q}_2,u_2}(x)\,\psi^+_{\Delta_3,\Tilde{q}_3,u_3}(y) \, .
	\end{split}
\end{equation}
Using the incoming and outgoing wavefunctions derived in eq.\eqref{eq:allpropagatorslarger}, we can simplify the integrand by performing integrations. One can show (see Appendix \ref{ap:threepointcalc} for details) that the integrand is proportional to
\begin{equation}\label{eq:threepointdelta}
	\delta^{(4)}(\omega_3 \Tilde{q}_3 - \omega_1\Tilde{q}_1 - \omega_2 \Tilde{q}_2) \, .
\end{equation}
Thus, we have
\begin{equation}
	\langle O_{\Delta_1}(\p_1)O_{\Delta_2}(\p_2)O_{\Delta_3}(\p_3) \rangle \simeq 0 \, ,
\end{equation}
for the two incoming-one outgoing diagram. This is because $\Tilde{q}_i \in \mathbb{R}^{1,3}$. One cannot have momentum conservation in $\mathbb{R}^{1,3}$ with just three momenta. One can see that from
\begin{equation}\label{eq:momcons}
	(\omega_1 \Tilde{q}_1 + \omega_2 \Tilde{q}_2)^2 = 2\omega_1 \omega_2 \Tilde{q}_1 \cdot \Tilde{q}_2 ~ \bm{\neq} ~ \omega^2_3 \Tilde{q}^2_3 \, .
\end{equation}
This means that the argument of the delta function \refb{eq:threepointdelta} is not satisfied and always gives zero. From symmetry, we can see that the diagram with one incoming particle exchanging an operator with two outgoing particles also evaluates to zero. 

For the contact diagram in figure \ref{fig:threepoint} also, we have a similar analysis. We have \footnote{We can consider all the possible cases of incoming and outgoing wavefunctions.}
\begin{equation}
	\langle O_{\Delta_1}(\p_1)O_{\Delta_2}(\p_2)O_{\Delta_3}(\p_3) \rangle = (i\mu)\int_{\text{AdS}_{4}}d^{4}\x \,\textbf{K}_{\Delta_{1}}(\p_1,\x)\textbf{K}_{\Delta_{3}}(\p_3,\x)\textbf{K}_{\Delta_{2}}(\p_2,\x) \, .
\end{equation}
Thus, in the $R \to \infty$ limit of eq.\eqref{eq:genwittenlarger}, we have
\begin{equation}\label{eq:threepointcontact}
	\begin{split}
		\langle O_{\Delta_1}(\p_1)O_{\Delta_2}(\p_2)O_{\Delta_3}(\p_3) \rangle &\simeq N^3_{\Delta_1}N^3_{\Delta_2}N^3_{\Delta_3} \int_{\mathbb{R}^{1,3}} d^{4}x \, \psi^{\pm}_{\Delta_1,\Tilde{q}_1,u_1}(x) \,\psi^{\pm}_{\Delta_2,\Tilde{q}_2,u_2}(x)\,\psi^{\pm}_{\Delta_3,\Tilde{q}_3,u_3}(x) \\
		& \propto \delta^{(4)}(\mp \omega_1 \Tilde{q}_1 \mp \omega_2 \Tilde{q}_2 \mp \omega_3 \Tilde{q}_3) = 0  \, .
	\end{split}
\end{equation}
We see that for generic momentum, the three point function evaluates to zero in the large $R$ limit. This matches with the expected result \cite{Banerjee:2018gce}, where we also have a vanishing three point function.

\subsubsection{Collinear limit}
\label{sec:threepointcollinear}
From eq.\eqref{eq:momcons}, we see that if $\Tilde{q}_1 =\Tilde{q}_2$, we can still have a non-zero result because now the three particles become collinear. Thus, the split of the momentum conserving delta function becomes
\begin{equation}\label{eq:momcons2}
	\delta^{4}(\omega_3 \Tilde{q}_3 -\omega_1\tilde{q}_1 -\omega_2\Tilde{q}_2) = \dfrac{1}{\omega^3_3}\delta(\omega_3 - \omega_1 - \omega_2)\delta(z_{12})\delta(z_{13})\delta(\bar{z}_{12})\delta(\bar{z}_{13}) \, .
\end{equation}
This particular momentum conserving delta function will ensure that eq.\eqref{eq:threepointfirst} is non-zero. Since, $\tilde{q}_1 = (1,\Omega_1)$, $z_1 = z_2 = z_3$ ensured by the delta functions in eq.\eqref{eq:momcons2} implies $\tilde{q}_1=\tilde{q}_2=\tilde{q}_3$ which corresponds to the collinear case. In fact, substituting eq.\eqref{eq:momcons2} in eq.\eqref{eq:threepointfirst}, we get the following result
\begin{equation}\label{eq:threepointcollinear}
	\begin{split}
	    \langle O_{\Delta_1}(\p_1)&O_{\Delta_2}(\p_2)O_{\Delta_3}(\p_3) \rangle \\ &= \mathcal{A}_{(3)}\delta^2(z_{12})\delta^2(z_{13}) \sum_{k=0}^{\Delta_3 -4} \dfrac{{}^{\Delta_3-4}C_k \, \Gamma(k+\Delta_1) \, \Gamma(\Delta_2+\Delta_3-k-4)}{(i(u_3-u_1))^{\Delta_1+k} \, (i(u_3-u_2))^{\Delta_2+\Delta_3-4-k}} \, .
	\end{split}
\end{equation}
Here $\mathcal{A}_{(3)}$ is a normalization constant analogous to eq.\eqref{eq:normconst}. The result is valid for $\Delta_3 \in \mathbb{N}$ and $\Delta_3 \geq 4$. A derivation of eq.\eqref{eq:threepointcollinear} is given in Appendix \ref{ap:threepointcollinear}. This result matches with the intrinsic Carrollian CFT result derived from the Ward identities of the Carrollian CFT in Sec.~3 (in particular eq.\eqref{eq:threepointpowers} and eq.\eqref{eq:threepointintrinsic}). 

\medskip

We emphasize that we don't really need the internal propagator to arrive at the results of eq.\eqref{result1} and eq.\eqref{eq:threepointcollinear}. If we had started from the contact diagram of eq.\eqref{eq:threepointcontact}, the final result would be the same apart from a factor of $\frac{1}{m^2}$. The imprint of the mass of the exchanged operator will show up in the normalization constants like $\mathcal{A}$ of eq.\eqref{eq:normconst}. This will not show up if you consider contact diagrams.

\subsubsection*{Some features}

We now illustrate certain interesting features of the result of eq.\eqref{eq:threepointcollinear}. For $\Delta_3 = 4$ in eq.\eqref{eq:threepointcollinear}, we have
\begin{equation}\label{eq:threepointcollineardelta31}
	\begin{split}
		\langle O_{\Delta_1}(\p_1)O_{\Delta_2}(\p_2)O_{4}(\p_3) \rangle &= \mathcal{A}_{(3)} \delta^2(z_{12})\delta^2(z_{13}) \dfrac{\Gamma(\Delta_1)\Gamma(\Delta_2)}{(i(u_3-u_1))^{\Delta_1} \, (i(u_3-u_2))^{\Delta_2}} \, \\
		& \propto \langle O_{\Delta_1}(\p_1) O_{2}(\p_3) \rangle \langle O_{\Delta_2}(\p_2) O_{2}(\p_3) \rangle \, .
	\end{split}
\end{equation}
Upto an overall normalization, this sort of special factorization in terms of the two point function of eq.\eqref{result1} seems to only work for $\Delta_3 =4$. To be precise, for $\Delta_3=4$, we have
\begin{equation}\label{eq:threepointfactordelta32}
	\begin{split}
	    \langle O_{\Delta_1}(\p_1)&O_{\Delta_2}(\p_2)O_{\Delta_3}(\p_3) \rangle \\&= \dfrac{m^2\,N^3_{\Delta_3}}{(2\pi)^{4}N^3_{\Delta_3-2}N^3_{\Delta_3-2}} \langle O_{\Delta_1}(\p_1) O_{\Delta_3-2}(\p_3) \rangle \langle O_{\Delta_2}(\p_2) O_{\Delta_3-2}(\p_3) \rangle \, .
	\end{split}
\end{equation}
For higher $\Delta_3$, we have a slightly more complicated factorization in terms of two point functions.

\medskip

For $\Delta_3 = 5$ in eq.\eqref{eq:threepointcollinear} we have
\begin{equation}
	\begin{split}
		\langle O_{\Delta_1}(\p_1)O_{\Delta_2}(\p_2)O_{5}(\p_3) \rangle &= \mathcal{A}_{(3)} \delta^2(z_{12})\delta^2(z_{13}) \left[ \dfrac{\Gamma(\Delta_1+1)\Gamma(\Delta_2)}{(i(u_3-u_1))^{\Delta_1+1} \, (i(u_3-u_2))^{\Delta_2}} \right. \\
		& \left. ~~~ + \dfrac{\Gamma(\Delta_1)\Gamma(\Delta_2+1)}{(i(u_3-u_1))^{\Delta_1} \, (i(u_3-u_2))^{\Delta_2+1}} \right] \, \\
		& \propto \left[ \langle O_{\Delta_1}(\p_1) O_{3}(\p_3) \rangle \langle O_{\Delta_2}(\p_2) O_{2}(\p_3) \rangle \right. \\
		& ~~~~ + \left. \langle O_{\Delta_1}(\p_1) O_{2}(\p_3) \rangle \langle O_{\Delta_2}(\p_2) O_{3}(\p_3) \rangle \right] \, .
	\end{split}
\end{equation}
This is more complicated than the factorization in eq.\eqref{eq:threepointfactordelta32} in that it involves two point correlators of lower $\Delta_3$ as well. One can see that the generic three point function given by eq.\eqref{eq:threepointcollinear} also has this factorization. To see it notice that
\begin{equation}\label{eq:twopointinthreepoint}
	\begin{split}
		\dfrac{\Gamma(k+\Delta_1)\delta^2(z_{31})}{(i(u_3-u_1))^{\Delta_1+k}} &= \dfrac{m^2}{(2\pi)^{4}N^3_{\Delta_1}N^3_{k+2}} \langle O_{\Delta_1}(\p_1) O_{k+2}(\p_3) \rangle \, , \\
		\dfrac{\Gamma(\Delta_2+\Delta_3-k-4)\delta^2(z_{32})}{(i(u_3-u_2))^{\Delta_2+\Delta_3-k-4}} &= \dfrac{m^2}{(2\pi)^{4}N^3_{\Delta_2}N^3_{\Delta_3-k-2}} \langle O_{\Delta_2}(\p_2) O_{\Delta_3-k-2}(\p_3) \rangle \, .
	\end{split}
\end{equation}
Since $\Omega_1 = \Omega_2=\Omega_3$, one can substitute eq.\eqref{eq:twopointinthreepoint} in eq.\eqref{eq:threepointcollinear} to obtain
\begin{equation}
	\begin{split}
		\langle O_{\Delta_1}(\p_1)O_{\Delta_2}(\p_2)O_{\Delta_3}(\p_3) \rangle = \dfrac{m^2(i\mu)}{(2\pi)^{4}} \sum_{k=0}^{\Delta_3-4} \dfrac{N^d_{\Delta_3} \, {}^{\Delta_3-4}C_k }{N^3_{k+2}N^3_{\Delta_3-k-2}} & \langle O_{\Delta_1}(\p_1) O_{k+2}(\p_3) \rangle \\ & ~~\times \langle O_{\Delta_2}(\p_2) O_{\Delta_3-k-2}(\p_3) \rangle \, .
	\end{split}
\end{equation}

Interestingly, for $u_2=u_1$ ($\p_1=\p_2$) in eq.\eqref{eq:threepointcollineardelta31} for $\Delta_3=4$ (the strict ``fully collinear limit"), we get
\begin{equation}
	\begin{split}
		\langle O_{\Delta_1}(\p_1)O_{\Delta_2}(\p_1)O_{\Delta_3}(\p_3) \rangle &= \mathcal{A}_{(3)} \delta^2(z_{12})\delta^2(z_{13}) \dfrac{\Gamma(\Delta_1)\Gamma(\Delta_2)}{(i(u_3-u_1))^{\Delta_1+\Delta_2} \, } \, \\
		& \propto \langle O_{\Delta_1+\Delta_2}(\p_1) O_{\Delta_3-2}(\p_3) \rangle \, .
	\end{split}
\end{equation}
The operator dimensions of $O_{\Delta_1},O_{\Delta_2}$ seem to add up resulting in an effective operator of dimension $\Delta_1+\Delta_2$. This structure is exactly the two point function derived in eq.\eqref{result1} upto a diverging $\delta^2(0)$. This structure is present in the generic three point function in the collinear limit as well. For instance, setting $u_1=u_2$ in  eq.\eqref{eq:threepointcollinear}, we have
\begin{equation}
	\begin{split}
		\langle O_{\Delta_1}(\p_1)&O_{\Delta_2}(\p_1)O_{\Delta_3}(\p_3) \rangle \\ &= \mathcal{A}_{(3)} \delta^2(z_{12})\delta^2(z_{13}) \dfrac{\sum_{k=0}^{\Delta_3-4} {}^{\Delta_3-4}C_k \Gamma(k+\Delta_1)\Gamma(\Delta_2+\Delta_3-k-4)}{(i(u_3-u_1))^{\Delta_1+\Delta_2+\Delta_3-4}} \\
		& \propto \langle O_{\Delta_1+\Delta_2}(\p_1)O_{\Delta_3-2}(\p_3) \rangle \, .
	\end{split}
\end{equation}
This completes our present analysis of the three point function from AdS Witten diagrams. 

\medskip

As a final point, we note that the bulk to boundary propagator is valid for particular Poincar\'e patches \cite{Penedones:2007ns}. We consider insertion points at $\mathcal{O}(R^{-1})$ regions about $\tau_p = \pm \frac{\pi}{2}$ and the scattering process occurs at the center of AdS. The regime of validity of the propagator inserted at $\tau_p = -\frac{\pi}{2}+\mathcal{O}(R^{-1})$ and the regime of validity of the propagator inserted at $\tau_p = +\frac{\pi}{2}+\mathcal{O}(R^{-1})$ overlap in a region that encompasses the center of the AdS. The subtlety of the validity of the bulk to boundary propagator of eq.\eqref{eq:bulkbdyprop} will not matter to us. With this we conclude our analysis of the large radius limit of AdS Witten diagrams. 
\newpage

\section{Discussions}

\subsection*{Our work in a nutshell}
In our work, we have built a prescription of obtaining correlation functions of a {\em{co-dimensional one}} holographic dual of asymptotically flat spacetimes by taking a large radius limit of AdS Witten diagrams and carefully keeping track of the null retarded (or advanced) time direction of the null boundary of the resulting flat spacetime. This is encapsulated in eq \refb{eq:genwittenlarger}. In particular we have shown that by taking this infinite radius limit on Witten diagrams in AdS$_4$, one can reproduce correlation functions of {\em{three dimensional}} field theory that is invariant under the (asymptotic) symmetries of 4d flat spacetime. This 3d dual field theory is a Carrollian CFT that lives on the entire null boundary and not just on the celestial sphere. By this method, we have reproduced (upto normalisations) the two and three point functions of the Carrollian CFT in the so-called delta function branch which we obtained first by using conformal Carroll Ward identities. Our construction generalises that of \cite{PipolodeGioia:2022exe}, the results of which we can simply obtain by putting the null coordinate to zero. As a by-product of our analysis, we have obtained an antipodal matching condition \refb{eq:antipodal} which is now valid for all retarded and advanced times and hence holds on the whole of the null boundary and not only on the celestial sphere. 

\subsection*{Importance}
When it was established that Carrollian CFT correlation functions connect to the one higher dimensional S-matrix elements in flat spacetimes via the modified Mellin transformations, it was unclear how or if at all this holographic correspondence was connected to AdS/CFT. The Carrollian CFT correlators in the delta function branch don't seem to have any direct connection to relativistic CFT correlation functions since e.g. the Carrollian 2-point function \refb{Sym-cor} can be non-zero even when the primary fields have unequal scaling dimensions and it is unclear how to relate this to the zero answer from the relativistic CFT. This hence was a potential stumbling block in understanding holography in asymptotically flat spacetimes from a large radius limit of AdS/CFT through the Carrollian perspective. Our work in this paper shows how to resolve this issue by working in embedding space and considering the null direction. 

\medskip

\subsection*{Future directions}

There are numerous directions of interesting work that would emerge from our explorations in this paper. Below we mention a few important ones. 

\medskip

Our prescription now provides a very general picture which now can be used to import well-known results from AdS/CFT to understand features of co-dimension one duals of asymptotically flat spacetimes. One of the immediate goals would be to understand the nature of four-point functions and what this means for Carrollian blocks and the conformal bootstrap programme in these field theories. While there has been work in understanding BMS blocks in 2d field theories \cite{Bagchi:2016geg, Bagchi:2017cpu}, it would be particularly intriguing to work out the subtleties associated with the delta-branch of the correlators. 

\medskip

Another direction of immediate interest is how to generalise this set up to include scattering of massive particles. While it seems that one can include scaling dimensions that are proportional to the AdS radius while taking the limit and possibly land up on massive scattering, the fact that all massive particles end up not on $\mathscr{I}^+$, but at $i^+$ may complicate matters. 

\medskip

While our procedure outlined in Sec.~3 is dimension independent, we have focussed on bulk dimensions $d=4$. In some sense, $d=3$ is more tractable and it would be good to analyse this in detail and compare and contrast to existing literature on the subject, including in particular \cite{Bagchi:2015wna}. 

\medskip

The methods outlined in our work can be looked upto a first step in an attempt to reformulate the bulk reconstruction programme of HKLL \cite{Hamilton:2005ju,Hamilton:2006az,Hamilton:2006fh} and adapt it to asymptotically flat spacetimes. This is of course a long-term goal. Again, there has been some work in the direction of bulk reconstruction for 3d Minkowski spacetimes \cite{Hartong:2015usd}. It would be good to revisit this in terms of a limit from AdS$_3$/CFT$_2$. 

\medskip

Finally, our paper gives some further insight as to how 3d Carrollian CFT and 2d Celestial CFTs may be related and it indicates that the lower dimensional Celestial CFT may be obtained as a restriction of a higher dimensional Carroll CFT to a constant $u$ slice. It would be good to make this connection concrete.



\bigskip \bigskip

\subsection*{Acknowledgements}
We thank Suvikrant Gera for initial collaboration and Shamik Banerjee, Rudranil Basu, Shankhadeep Chakrabortty, Diptarka Das, Shahin Sheikh Jabbari, Kedar Kolekar, Arthur Lipstein, Amartya Saha and Joan Simon for interesting discussions and comments. PD would like to thank Milan Patra and Anurag Sarkar for valuable discussions.

\smallskip

The work of AB is partially supported by a Swarnajayanti fellowship (SB/SJF/2019-20/08) from the Science and Engineering Research Board (SERB) India, the SERB grant (CRG/2020/002035), a visiting professorship at \'{E}cole Polytechnique Paris, and a Royal Society of London international exchange grant with the University of Edinburgh. AB also acknowledges the warm hospitality of the Niels Bohr Institute, Copenhagen and the University of Edinburgh, UK during various stages of this work. PD would like to duly acknowledge the Council of Scientific and Industrial
Research (CSIR), New Delhi for financial assistance through the Senior Research Fellowship (SRF) scheme. SD would like to thank department of physics, IIT Kanpur for partial financial support.

\newpage

\section*{Appendices}

\appendix

\section{Modified Mellin transformations}
\label{ap:modifiedmellin}

\subsection{Mellin transformations and Celestial Primaries}
The basic observables in asymptotically flat spacetimes are the scattering amplitudes.  Celestial CFT emerges as dual description of gravity theories in asymptotically flat spacetimes as it can successfully encapsulate the scattering amplitudes in the bulk. According to this proposal the dual field theory is a 2d CFT that lives on the celestial sphere of null infinity. The holographic correspondence between the S-matrix in flat spacetime and the dual celestial CFT is given by \cite{Pasterski:2017kqt}: 
\begin{align} \label{scattering}
\<\mathcal{O}_{1}(z_1,\bar{z}_1)\mathcal{O}_2(z_2,\bar{z}_2)...\mathcal{O}_n(z_n,\bar{z}_n)\>\sim \int_{0}^{\infty}\prod_{i=1}^n \omega_{i}^{\Delta_i-1}\mathcal{S}_n(\omega_i,z_i,\bar{z}_i,\sigma_i)
\end{align}
On the left hand side we have the correlation functions of the primary operators $\mathcal{O}(z,\bar{z})$ on celestial sphere and on the right we have Mellin transformations of  $\mathcal{S}_n$, which denotes the $n$-point scattering amplitudes of massless particles in the bulk.  

\medskip

Let's discuss about this formula in a little more detail.  We are interested in scattering of massless particles. A suitable parametrisation of the four momentum of a massless particle is
\begin{equation}
p^{\mu}=\omega \Big( 1+z\bar{z},z+\bar{z},i(z-\bar{z}),1-z\bar{z}\Big)
\end{equation}
Here $\omega$ carries the information about the  energy of the particle and $z$ and $\bar{z}$ represents the stereographic coordinates on the celestial sphere. The action of Lorentz group in this parametrisation is manifest as the group of global conformal transformations on celestial sphere. The double cover of the Lorentz group, SL(2,C) acts on the particle momenta  $p^\mu$ as 
 \begin{align} \label{Lorentz}
  z\to z'=\frac{az+b}{cz+d},  \quad \bar{z}\to \bar{z}'= \frac{\bar{a}\bar{z}+\bar{b}}{\bar{c}\bar{z}+\bar{d}}, \quad \omega \to \omega'= \omega (cz+d)(\bar{c}\bar{z}+\bar{d})\omega
 \end{align}

The single particle states in the bulk can be obtained  by the Fock space creation  operators $a(p^\mu,\sigma)$ / $a^\dag(p^\mu,\sigma)$. For example a state with momentum $p^\mu$ and helicity $\sigma$ is created from the vacuum  by 
\begin{align}
|p^\mu,\sigma\rangle=a ^\dag(p^\mu,\sigma) |0 \rangle
\end{align}
In compact notation we shall write these as $a(\epsilon\omega,z,\bar{z},\sigma)$, where $\epsilon=\pm1$ decides if this a creation or an annihilation operator. Using these operators in the bulk we can define a local field $\mathcal{O}(z,\bar{z})$  in the following way:
\begin{align} \label{mellin}
\mathcal{O}_{h,\bar{h}}(z,\bar{z})=\int_{0}^{\infty}d\omega \omega^{\Delta-1} a(\epsilon \omega,z,\bar{z},\sigma).
\end{align}
The weights $h$ and $\bar{h}$ of these operators can be determined in terms of $\Delta$ and spin $\sigma$ as
 \begin{align}
 h+\bar{h}=\Delta \quad \text{and} \quad  h-\bar{h}=\sigma.
 \end{align}
Now as the scattering amplitudes in the bulk can be viewed as the correlation function of these creation and annihilation operators, they can be recast into the correlators of these local fields $\mathcal{O}(z,\bar{z})$ on celestial sphere giving rise to (\ref{scattering}). 
 
\medskip

Transformation rules of these operators under the bulk Poincare group can also be figured out directly using (\ref{mellin}). The Lorentz group acts on celestial sphere as a group of global conformal transformations.
$\mathcal{O}(z,\bar{z})$  can  transforms like a SL(2,$\mathbb{C}$) primary field with conformal weights  $h$ and spin $\bar{h}$. i.e, 
\begin{align}
\mathcal{O}_{h,\bar{h}}(z,\bar{z}) \to \mathcal{O}_{h,\bar{h}}(z',\bar{z}')=\frac{1}{(cz+d)^{2h}}\frac{1}{(\bar{c}\bar{z}+\bar{d})^{2\bar{h}}} \mathcal{O}_{h,\bar{h}}(z,\bar{z})
\end{align}
However the action of the bulk translations $p^\mu \to p^\mu+l^{\mu}$ on these primaries is non-trivial. Consider for example the time translation in the bulk generated by the bulk Hamiltonian $H$. The celestial primary field transforms by $H$ in the following way
\begin{align}\label{celH}
\delta_H \mathcal{O}(z,\bar{z})&=\int_0^{\infty} d\omega \omega^{\Delta-1}[H,a(\epsilon\omega,z,\bar{z},\sigma)]  \\ \nonumber
&= \int_0^{\infty} d\omega \omega^{\Delta-1} [-\epsilon\omega(1+z\bar{z}) a(\epsilon\omega,z,\bar{z},\sigma)] \\ \nonumber
&=-\epsilon (1+z\bar{z}) \mathcal{O}_{h+\frac{1}{2},\bar{h}+\frac{1}{2}}(z,\bar{z})
\end{align}
The action of the other translation generators are similar. One obvious peculiarity in this relation is the shift of the holomorphic and anti-holomorphic weights of these celestial primaries. The other relations are given by
\begin{align}\label{celP}
[P^1,\mathcal{O}(z,\bar{z})]&=-\epsilon (z+\bar{z})\mathcal{O}_{h+\frac{1}{2},\bar{h}+\frac{1}{2}}(z,\bar{z}),  \\ \nonumber
 [P^2,\mathcal{O}(z,\bar{z})]&=i\epsilon(z-\bar{z})\mathcal{O}_{h+\frac{1}{2},\bar{h}+\frac{1}{2}}(z,\bar{z})  \\ \nonumber
 [P^3,\mathcal{O}(z,\bar{z})]&=-\epsilon(1-z\bar{z})\mathcal{O}_{h+\frac{1}{2},\bar{h}+\frac{1}{2}}(z,\bar{z}) 
\end{align}
We believe this changing of operator weights under spacetime translations is not a desirable feature and will now discuss a natural prescription to fix this. 

\subsection{Carroll and Modified Mellin}
Although celestial CFTs have been very successful in capturing the scattering amplitudes and their infrared behaviour, our motivation is to describe the flat spacetimes holographically in terms of a co-dimension one field theories that live on the whole of the null boundary of flat spacetimes. To us, this is more natural because it is possible to understand the connection of flat holography with AdS/CFT by taking a suitable zero cosmological limit and this does not reduce the dimension of the dual field theory. These so called Carrollian CFTs on the null boundary have also shown to be able to describe the scattering processes in asymptotically flat spacetimes \cite{Bagchi:2022emh}. To understand the possible relation of these Carroll CFTs and celestial CFT, we discuss the `Heisenberg picture' of the above construction. 

\medskip

Let's consider the evolution of $\mathcal{O}(z,\bar{z})$ by the Hamiltonian $H$
\begin{align}
\Phi^{\epsilon}_{h,\bar{h}}(u,z,\bar{z})&=e^{-iHU}\mathcal{O}_{h,\bar{h}}(z,\bar{z})e^{iHU} =\int_{0}^{\infty} d\omega \, \, \omega^{\Delta-1}e^{-iHU}a(\epsilon\omega,z,\bar{z})e^{iHU} \\ \nonumber
&=\int_{0}^{\infty}d\omega \, \, \omega^{\Delta-1}e^{-i\epsilon\omega u}a(\epsilon\omega,z,\bar{z}).
\end{align}
Here we have rescaled $U$ and have defined $u=U(1+z\bar{z})$. In the last line we have obtained the {\em{modified Mellin transformation}} \cite{Banerjee:2018gce}. This newly defined field $\Phi^{\epsilon}(u,z,\bar{z})$ lives on a 3d space spanned by $u,z$ and $\bar{z}$. We can confirm this is the null boundary by deriving the transformation rules of this field $\Phi^{\epsilon}(u,z,\bar{z})$. In similar lines to celestial CFT, we can show under Lorentz transformations given by (\ref{Lorentz}), $\Phi^{\epsilon}(u,z,\bar{z})$ transforms as \cite{Banerjee:2018gce}
 \begin{align}
 \Phi^{\epsilon}_{h,\bar{h}}(u,z,\bar{z}) \to \Phi^{\epsilon}_{h,\bar{h}}(u',z',\bar{z}')=\frac{1}{(cz+d)^{2h}}\frac{1}{(\bar{c}\bar{z}+\bar{d})^{2\bar{h}}}
 \Phi^{\epsilon}\Big(\frac{u}{|cz+d|^2},\frac{az+b}{cz+d},\frac{\bar{a}\bar{z}+\bar{b}}{\bar{c}\bar{z}+\bar{d}}\Big)
 \end{align}
Infinitesimally this transformation is same as (\ref{Primary transformations}) for $n=0,\pm1$. This is the action of Lorentz group on the null boundary ($\mathscr{I}^\pm$).

\medskip

Now, the action of the bulk translations are given by \cite{Banerjee:2018gce}
\begin{align}
 \Phi^{\epsilon}_{h,\bar{h}}(u,z,\bar{z}) \to \Phi^{\epsilon}_{h,\bar{h}}(u',z',\bar{z}')=\Phi^{\epsilon}_{h,\bar{h}}(u+p+qz+r\bar{z}+sz\bar{z},z,\bar{z})
\end{align}
These transformations are again equivalent to (\ref{Primary transformations}) for $r,s=0,1$. The noteworthy difference here with the previous transformations of the Celestial primary operators \refb{celH}, \refb{celP} is that the bulk translations don't shift the weights any more. This is because of the presence of $u$ co-ordinate and this makes the operators transform more naturally \cite{Banerjee:2018gce}. These transformation properties certify that the above construction gives rise to a field on $\mathscr{I}^\pm$.

\medskip

Having this construction at hand one would be able to relate the scattering amplitudes in asymptotically flat spacetimes and the correlation function of a co-dimension one field theory that lives on the whole of the null boundary $\mathscr{I}^\pm$.  This formula is given by
\begin{align} \label{Scattering new}
\langle \Phi_1^{\epsilon_1}(u_1,z_1,\bar{z}_1)\Phi_2^{\epsilon_2}(u_2,z_2,\bar{z}_2).... \Phi_n^{\epsilon_n}(u_n,z_n,\bar{z}_n)\rangle=\int_{0}^{\infty}\prod_{i=1}^{n} d\omega_i e^{-i\epsilon_i\omega_iu_i}\omega_i^{\Delta_i-1}\mathcal{S}_n
\end{align}
On the one hand, from a purely flat space point of view, this makes the Mellin amplitudes finite \cite{Banerjee:2019prz}. On the other, we now have a way to relate correlation functions on the co-dimension one dual Carrollian CFT to scattering amplitudes in the bulk.

\subsection{Two point correlator}
Now let us consider the simplest case of free propagating amplitudes. If a massless particle starts off at $\mathscr{I}^{-}$ and arrives at $\mathscr{I}^{+}$ without any interaction in the deep interior. Then this amplitude is given by
\begin{align}
\langle p_1,\sigma_1|p_2,\sigma_2\rangle=&(2\pi)^32E_{p_1}\delta^3(p_1-p_2)\delta_{\sigma_1+\sigma_2} = 4\pi^3 \frac{\delta(\omega_1-\omega_2)}{\omega_1}\delta^{(2)}(z_1-z_2)\delta_{\sigma_1+\sigma_2,0}
\end{align}
If we just take this expression and plug it in the modified Mellin transformation of S-matrix elements (\ref{Scattering new}), we should expect to obtain the two point correlation function of 3d Carrollian CFT. So we do this and get:
\begin{equation}
	\begin{split}
		\langle &\Phi_1(u_1,z_1,\bar{z}_1)\Phi_2(u_2,z_2,\bar{z}_2)\> \\
		&\hspace{1.5cm} =4\pi^3\delta_{\sigma_1+\sigma_2,0}\int_{0}^{\infty} d\omega_1 d\omega_2 \omega_1^{\Delta_1-1}\omega_2^{\Delta_2-1}e^{-i\omega_1u_1}e^{i\omega_2u_2}\frac{\delta(\omega_1-\omega_2)}{\omega_1}\delta^{(2)}(z_1-z_2)
	\end{split}
\end{equation} 
\begin{align}
\Rightarrow \quad &\langle \Phi_1(u_1,z_1,\bar{z}_1)\Phi_2(u_2,z_2,\bar{z}_2)\> =4 \pi^3 \Gamma(\Delta_1+\Delta_2-2) \frac{\delta^2(z_1-z_2)}{i(u_1-u_2)^{\Delta_1+\Delta_2-2}}\delta_{\sigma_1+\sigma_2,0}
\end{align}
This matches up with the two point function associated with the delta function branch that we derived from the symmetry arguments in (\ref{Sym-cor}), thereby providing further evidence that this co-dimension one holographic correspondence between asymptotically flat spacetimes and Carroll CFTs living on the entire null boundary holds.

\section{A review of the embedding space representation}
\label{ap:embedding}

In this Appendix, we briefly review the elements of the embedding space representation that we will need in our analysis. For details, we refer the reader to \cite{Penedones:2007ns}.

\medskip

$\text{AdS}_{d+1}$ can be defined in the embedding space $\mathbb{R}^{2,d}$ as the set of points $\x$:
\begin{equation}
	\x \in \mathbb{R}^{2,d}, \hspace{2cm} \x^2 = - R^2
\end{equation}
The holographic boundary of $\text{AdS}_{d+1}$ is defined as the set of outward null rays $\p$ from the origin of the embedding space $\mathbb{R}^{2,d}$:
\begin{equation}
	\p \in \mathbb{R}^{2,d}, \hspace{0.7cm} \p^2 = 0, \hspace{0.7cm} \p \sim \lambda \p ~ (\lambda > 0)
\end{equation}
These null rays form a light cone. Usually, the AdS boundary is identified as a specific section $\Sigma$ of the light cone. The particular example of a $d$-dimensional AdS  solution that we work with is elaborated below eq.\eqref{embedding} in section \ref{sec:wittendiagram}. For the purposes of this section, let us set $R=1$. The factors of $R$ can easily be reinstated.

\medskip

The Laplacian on $\text{AdS}_{d+1}$ takes a simple form in the embedding space:
\begin{equation}\label{eq:boxads}
	\Box_{\text{AdS}_{d+1}} = \bm{\partial}^2 + \x \cdot \bm{\partial}(d + \x \cdot \bm{\partial}) \, .
\end{equation}
This implies that if a scalar field satisfies the massless Klein Gordon equation in the embedding space,
\begin{equation}
	\bm{\partial}^2 \psi(\x) = 0 \, ,
\end{equation}
and obeys the scaling relation
\begin{equation}
	\psi(\lambda \, \x) = \lambda^{-\Delta} \, \psi(\x) \, ,
\end{equation}
then, it obeys the massive Klein Gordon equation with mass given by $m^2 = \Delta (\Delta -d)$ in AdS.

The Feynmann propagator in $\text{AdS}_{d+1}$ (with the appropriate $i\epsilon$ prescription as $\x$ crosses the lightcone of $\Bar{\x}$) is given by
\begin{equation}
	[\Box_{\text{AdS}_{d+1}} - \Delta (\Delta -d) ] \Pi_{\Delta}(\x,\Bar{\x}) = i \, \delta(\x,\Bar{\x}) \, .
\end{equation}
This can be solved by Wick rotating to the Hyperbolic space $\text{H}_{d+1}$ which is the Euclidean version of $\text{AdS}_{d+1}$. The Wick rotation results in
\begin{equation}
	[\Box_{\text{H}_{d+1}} - \Delta (\Delta -d) ] \Pi_{\Delta}(\x,\Bar{\x}) = - \, \delta(\x,\Bar{\x}) \, .
\end{equation}
The propagator  is thus given by \cite{BURGESS1985137}
\begin{equation}\label{eq:bkbkprop}
	\Pi_{\Delta} = C^d_{\Delta} \, a^{-\Delta} \, _2F_1 \left(\Delta, \dfrac{2\Delta - d +1}{2}, 2\Delta -d +1; -\dfrac{4}{a}\right) \, ,
\end{equation}
where
\begin{equation}
	\begin{split}
		C^d_{\Delta} &= \dfrac{1}{2 \pi^{\frac{d}{2}}} \dfrac{\Gamma(\Delta)}{\Gamma(\Delta - \dfrac{d}{2}+1)} \\
		a &= (\x - \Bar{\x})^2 = \x^2 + \Bar{\x}^2 - 2 \x \cdot \Bar{\x} = -2 (1+ \x \cdot \Bar{\x}) \, .
	\end{split}
\end{equation}
$a$ is the Lorentz invariant chordal distance which is always positive in the Euclidean regime. To find the factors of $R$, we resort to dimensional analysis. The corresponding equation is
\begin{equation}
    \left[\Box_{\text{H}_{d+1}} - \dfrac{\Delta (\Delta -d)}{R^2} \right] \Pi_{\Delta}(\x,\Bar{\x}) = - \, \delta(\x,\Bar{\x}) \, .
\end{equation}
Since $\x^2 = -R^2$, we should rescale the $\x$ in the above equations as $\x \to R \, \x$. The Hypergeometric function is just a number and thus is dimensionless. The dimensions of $R$ should come from $C^d_{\Delta}$ and $a^{-\Delta}$. Here $a \to R^2 a$. Thus,
\begin{equation}
     \delta(\x,\Bar{\x}) \to  \dfrac{1}{R^{d+1}}\delta(\x,\Bar{\x}) \implies C^d_{\Delta} \to \dfrac{1}{R^{d-1-2\Delta}}C^d_{\Delta} \, .
\end{equation}
This factor of $R$ should be compared with eq.\eqref{eq:cddelta}. This comes from the fact that
\begin{equation}
    \text{Dim}\left[\dfrac{1}{R^2} \Pi_{\Delta}(\x,\bar{\x}) \right] = \text{Dim}[\delta(\x,\Bar{\x})] \, .
\end{equation}

We will now obtain the bulk to boundary propagator from eq.\eqref{eq:bkbkprop} by letting one of the points go to the boundary. To reach the boundary of the embedding space coordinates, we first introduce generic coordinates of the form
\begin{equation}\label{eq:genericbdycoord}
	\x = \dfrac{1}{y}\mathbf{p}(\mathbf{y}) + y \,\Bar{\mathbf{p}}(\mathbf{y}) \, .
\end{equation}
The metric in these coordinates is given by
\begin{equation}
	\begin{split}
		ds^2 &= d\x . d\x = \left( -\dfrac{1}{y^2}\p(\mathbf{y})dy + \dfrac{1}{y}d\p + \Bar{\p}(\mathbf{y})dy + y d\Bar{\p} \right)^2 \\
		&= \dfrac{1}{y^2} dy^2 + \dfrac{1}{y^2}d\p^2 + \dfrac{2}{y}\Bar{\p}\cdot d\p \, dy - \dfrac{2}{y}\p\cdot d\Bar{\p} \, dy + d\p \cdot d\Bar{\p} + y^2 d\Bar{\p}^2 \\
		&=\dfrac{1}{y^2}(dy^2 + ds^2_{\Sigma} + \mathcal{O}(y)) \, .
	\end{split}
\end{equation}
Thus, the boundary of embedding space coordinates is reached by using eq.\eqref{eq:genericbdycoord} in the $y \to 0$ limit.

The bulk to boundary propagator can thus be obtained as a limit of the bulk to bulk propagator.
\begin{equation}
	\textbf{K}_{\Delta} (\p,\x) = \lim_{y \to 0} y^{-\Delta} \, \Pi_{\Delta} \left(\dfrac{1}{y}\p + y \, \Bar{\p},\x \right) \, .
\end{equation}
Now as $y \to 0$, we have
\begin{equation}
	\begin{split}
		a &= -2 - 2\left( \dfrac{1}{y}\p + \Bar{\p} y \right) \cdot \x \rightarrow - \dfrac{2}{y} \p \cdot \x \, , \\
		& _2F_1 \left(\Delta, \dfrac{2\Delta - d +1}{2}, 2\Delta -d +1; \dfrac{2y}{\p\cdot\x}\right) \rightarrow 1 \, .
	\end{split}
\end{equation}
Thus, we finally get
\begin{equation}\label{eq:bbdpmain}
	\textbf{K}_{\Delta}(\p,\x) = \lim_{y \to 0} y^{-\Delta} \dfrac{C^d_{\Delta}}{\left(\dfrac{-2\p\cdot\x}{y}\right)^{\Delta}} = \dfrac{C^d_{\Delta}}{(-2 \p\cdot\x + i \epsilon)^{\Delta}} \, .
\end{equation}
This constitutes a derivation of the bulk to boundary propagator eq.\eqref{eq:bulkbdyprop}. One can straightforwardly see that it satisfies the massive Klein Gordon equation using the following:
\begin{equation}
	\begin{split}
		\x\cdot\bm{\partial} \, \textbf{K}_{\Delta} (\p,\x) = \dfrac{C^d_{\Delta}(-\Delta)(-2\p\cdot\x)}{(-2\p\cdot\x + i \epsilon)^{\Delta + 1}} = - \Delta \, K_{\Delta} (\p,\x) \, , \\
		\bm{\partial^2} \textbf{K}_{\Delta}(\p,\x) = \dfrac{C^d_{\Delta}(-\Delta)(-\Delta-1)(-2\p)\cdot(-2\p)}{(-2\p\cdot\x)^{\Delta + 2}} = 0 ~~~ \text{as} ~~~ \p^2 = 0 \, .
	\end{split}
\end{equation}

\section{Witten diagrams: Details of the computations}
In this appendix, we will collect the detailed calculations. We will flesh out the details of the two point function and three point function Witten diagram computation of sections \ref{sec:twopoint} and \ref{sec:threepoint} respectively.

\subsection{Bulk to boundary propagator in the large R limit}
\label{ap:bbdplarger}

In this section, we will derive eq.\eqref{eq:bbdplargerexp} from eq.\eqref{eq:bulkbdyprop}. Our embedding space coordinates are given by eq.\eqref{eq:embeddingcoord}. The corresponding boundary coordinates are given by eq.\eqref{eq:boundarycoord}:
\begin{equation}
    \begin{split}
        P^+ &= -\dfrac{1}{2}(\cos\,\tau_p - \Omega^4_p) \, , \hspace{2cm} P^- = -\dfrac{1}{2}(\cos\,\tau_p + \Omega^4_p)\, , \\
        P^1 &= -\dfrac{1}{2}\sin\,\tau_p \, , \hspace{3.23cm} P^i = \dfrac{1}{2}\Omega_i \, .
    \end{split}
\end{equation}
In the embedding space language, the boundary points satisfy $\p^2=0$. The norm uses the embedding space metric eq.\eqref{eq:embedmetric}. We can see this in the following:
\begin{equation}
    \begin{split}
        \p^2 &= -P^+ P^- - (P^1)^2 + (P^2)^2 + (P^3)^2 \\
        &= -\dfrac{1}{4}(\cos^2\tau_p - (\Omega^4_p)^2) - \dfrac{1}{4}\sin^2\tau_p + \dfrac{1}{4}(\Omega^2)^2 + \dfrac{1}{4}(\Omega^3)^2 = 0 \, .
    \end{split}
\end{equation}
In the large $R$ limit, we have $\tau = \dfrac{t}{R}$, \,$\rho = \dfrac{r}{R}$. Substituting in eq.\eqref{eq:embeddingcoord}, we have
\begin{equation}
    \begin{split}
        X^+ &= -R\left(1 - \dfrac{r}{R}\Omega^4 \right) \, , \hspace{2cm} X^- =-R \left(1 + \dfrac{r}{R}\Omega^4 \right)\, ,\\
 	X^1&=-t \, , \hspace{4.17cm}  X^i = r\, \Omega_i\, .
    \end{split}
\end{equation}
Thus, we finally have 
\begin{equation}
    \begin{split}
        \p\cdot \x &= -\dfrac{1}{2}P^+X^- -\dfrac{1}{2}P^-X^+ - P^1 X^1 + P^2 X^2 + P^3 X^3 \\
        &= -\dfrac{1}{2}R \, \cos\,\tau_p - \dfrac{1}{2}t\,\sin\,\tau_p + \dfrac{1}{2} r \, \Omega_p \cdot \Omega \, .
    \end{split}
\end{equation}
Substituting the above equation in eq.\eqref{eq:bulkbdyprop}, we straightforwardly get eq.\eqref{eq:bbdplargerexp}.

\subsection{Two point function calculation}
\label{ap:twopointcalc}
In this section, we aim to derive the result of eq.\eqref{result1}. We start with eq.\eqref{eq:twopointfirst} given by
\begin{equation}
	\begin{split}
		\langle O_{\Delta_1}(\p_1)O_{\Delta_2}(\p_2)\rangle=&  N_{\Delta_1}^{3}N_{\Delta_2}^{3}\int_{\mathbb{R}^{1,3}}d^{4}x \, d^{4}y \, d\omega_1 \, d\omega_2 \, d^{4}k \, \omega_1^{\Delta_1-1}\omega_2^{\Delta_2-1}e^{-\epsilon(\omega_1+\omega_2)}\\ &\times e^{i\omega_1. u_1} e^{-i\omega_2. u_2} \frac{e^{i(\omega_1\tilde{q}_1+k)\cdot x} e^{-i(\omega_2\tilde{q}_2+k)\cdot y}}{k^2+m^2+i\epsilon}\dfrac{1}{(2\pi)^{4}} \, .
	\end{split}
\end{equation}
Now
\begin{equation}\label{eq:twopointinbet1}
	\begin{split}
		&\int d^{4}x \,  d^{4}y \, d^{4}k \, \dfrac{e^{i(\omega_1 \Tilde{q}_1 + k) \cdot x}e^{-i(\omega_2\Tilde{q}_2 + k)\cdot y}}{k^2 + m^2 + i \epsilon} \\
		&~= (2\pi)^{4}\int d^{4}y \, d^{4}k \, \dfrac{\delta^{4}(\omega_1 \Tilde{q}_1 + k)e^{-i(\omega_2\Tilde{q}_2+k)\cdot y}}{k^2+m^2+i\epsilon} \\
		&~= (2\pi)^{4} \int d^{4}y \dfrac{e^{-i(\omega_2\Tilde{q}_2 - \omega_1 \Tilde{q}_1)\cdot y}}{\omega^2_1\Tilde{q}^2_1+m^2+i\epsilon} \\
		&~= (2\pi)^{8}\dfrac{\delta^{(4)}(\omega_2 \Tilde{q}_2 - \omega_1 \Tilde{q}_1)}{m^2} \, ,
	\end{split}
\end{equation}
where we have been able to write the final step because $\Tilde{q}^2_1=0$. We note that since $\tilde{q}_1$ corresponds to ingoing momenta, $\Tilde{q}_1 = (1,-\Omega_1)$ because of the antipodal matching condition of eq.\eqref{eq:antipodal} with the  parametrization given by eq.\eqref{eq:4dparamq}. Now comes the crucial point. In the delta function above, we must compare the directions in the sphere at one time slice. The direction of $\tilde{q}_2$ is with respect to the sphere at $\tau_p = +\frac{\pi}{2}+\frac{u_2}{R}$. The direction of $\tilde{q}_1$ is however, with respect to the sphere at $\tau_p = -\frac{\pi}{2}+\frac{u_1}{R}$. We use the antipodal matching condition of eq.\eqref{eq:antipodal} to express $\tilde{q}_1$ as $\tilde{q}_1 = (1,\Omega_1)$ where now the direction is with respect to the sphere at $\tau_p = +\frac{\pi}{2}+\frac{u_1}{R}$. If we set $u=0$, we are essentially comparing directions in one sphere at $\tau_p = +\frac{\pi}{2}$. In the parametrization of eq.\eqref{eq:4dparamq}, the antipodal point is reached by $z \to -\frac{1}{\bar{z}}$. Hence we have
\begin{equation}
	\tilde{q}^{\mu}_1 = (1,-\Omega_1) \xrightarrow{z \to -\frac{1}{\bar{z}}} (1,\Omega_1) = \left[1,\dfrac{z_1+\bar{z}_1}{1+z_1\bar{z}_1},\dfrac{-i(z_1-\bar{z}_1)}{1+z_1\bar{z}_1},\dfrac{1-z_1\bar{z}_1}{1+ z_1\bar{z}_1} \right] \, .
\end{equation}
If we didn't have the antipodal matching condition, we would have gotten the unphysical
\begin{equation}
    \delta^{(4)}(\omega_2 \tilde{q}_2 - \omega_1 \tilde{q}_1) \propto \delta^{(3)}(\Omega_2 + \Omega_1) \propto \delta(z_2 + z_1)\delta(\bar{z}_2 + \bar{z}_1) \, .
\end{equation}
This will not satisfy the translation invariance. The antipodal matching condition eq.\eqref{eq:antipodal} ensures that crucial minus signs are present. Comparing the directions at the sphere at $\tau = \frac{\pi}{2}+\frac{u}{R}$ essentially means we are looking at the scattering amplitude from the perspective of future null infinity $\mathscr{I}^+$ only.

Thus, we can split the $3+1$ dimensional delta function to simplify the final step of eq.\eqref{eq:twopointinbet1} as
\begin{equation}\label{eq:twopointdelta}
	\delta^{(4)}(\omega_2 \tilde{q}^{\mu}_2-\omega_1 \tilde{q}^{\mu}_1) = \dfrac{1}{\omega_1}\delta(\omega_2-\omega_1)\delta(z_2-z_1)\delta(\bar{z}_2-\bar{z}_1) \, .
\end{equation}
The RHS is effectively a three dimensional delta function because of the null condition $\tilde{q}^2=0$. This combination is the lorentz invariant combination as can be seen from the analysis of Appendix \ref{ap:lorentzdelta}. Thus, eq.\eqref{eq:twopointinbet1} evaluates to 
\begin{equation}
	(2\pi)^{8} \dfrac{\delta(\omega_2 - \omega_1)\delta^2(z_2 - z_1)}{\omega_1 m^2 + i\epsilon} \, ,
\end{equation}
where $\delta^2(z_2-z_1) = \delta(z_{21})\delta(\bar{z}_{21})$. Substituting this result in eq.\eqref{eq:twopointfirst}, we have
\begin{align}\label{eq:twopointinbet2}
	\langle O_{\Delta_1}(\p_1)O_{\Delta_2}(\p_2)\rangle=& (2\pi)^{4}N_{\Delta_1}^{3}N_{\Delta_2}^{3}\int d\omega_1 \, d\omega_2 \, \omega_1^{\Delta_1-1}\,\omega_2^{\Delta_2-1}\,e^{-\epsilon(\omega_1+\omega_2)}\\ &\times e^{i\omega_1. u_1} e^{-i\omega_2. u_2} \frac{\delta(\omega_2-\omega_1)\delta^2(z_2-z_1) }{\omega_1 m^2+i\epsilon}\nonumber\\
	=& (2\pi)^{4}N_{\Delta_1}^{3}N_{\Delta_2}^{3}\int d\omega \, \omega^{\Delta_1+\Delta_2-3}e^{-2\epsilon\omega} e^{i\omega (u_1-u_2)} \delta^2(z_2-z_1) \, . \nonumber
\end{align}
One can evaluate the final $\omega$ integral using the following expression
\begin{equation}\label{eq:omegaint}
	\int d\omega \, \omega^{\Delta - k} e^{i\omega u} = (-iu)^{-1+k-\Delta} \Gamma(1-k+\Delta) \hspace{0.7cm} \text{if}~~~\text{Re}(k-\Delta)<1\, , ~\text{Im}(u)>0 \, .
\end{equation}
Crucially, we have been able to use this formula because of the $e^{-2\epsilon \omega}$ term that was present in eq.\eqref{eq:twopointinbet2} due to the $i\epsilon$ prescription. This makes the integrals like these convergent. Thus, using eq.\eqref{eq:omegaint} in eq.\eqref{eq:twopointinbet2}, we obtain the main result of our paper given by eq.\eqref{result1}:
\begin{equation}
	\langle O_{\Delta_1}(\p_1)O_{\Delta_2}(\p_2)\rangle = \mathcal{A} \frac{\delta^2(z_2-z_1)}{(i(u_2-u_1))^{\Delta_1+\Delta_2-2}} \, .
\end{equation}

\subsection{Three point function calculation}
We will outline the calculations concerning the three point function analyzed in section \ref{sec:threepoint}.

\subsubsection{Generic momenta}
\label{ap:threepointcalc}

For the case of generic ingoing and outgoing momenta, we first show eq.\eqref{eq:threepointdelta}. If we substitute eq.\eqref{eq:allpropagatorslarger} into eq.\eqref{eq:threepointfirst}, we have 
\begin{equation}\label{eq:threepointflatsecond}
	\begin{split}
		\langle O_{\Delta_1}(\p_1)O_{\Delta_2}(\p_2)O_{\Delta_3}(\p_3) \rangle \simeq 
		(i\mu)N^3_{\Delta_1}N^3_{\Delta_2}N^3_{\Delta_3} \int_{\mathbb{R}^{1,3}}d^{4}x \, d^{4}y \, d\omega_1 \, d\omega_2 \, d\omega_3 \, d^{4}k \\
		\times \omega^{\Delta_1 -1}_1 \omega^{\Delta_2-1}_2 \omega^{\Delta_3-1}_3 e^{i\omega_1 u_1}e^{i\omega_2 u_2} e^{-i\omega_3 u_3} e^{-\epsilon (\omega_1+\omega_2+\omega_3)} \\
		\times \dfrac{e^{i(\omega_1 \Tilde{q}_1 + \omega_2 \Tilde{q}_2 +k)\cdot x}e^{-i(\omega_3 \Tilde{q}_3 + k)\cdot y}}{k^2+m^2+i\epsilon}\dfrac{1}{(2\pi)^{4}} \, .
	\end{split}
\end{equation}
Consider a part of the integrand
\begin{equation}\label{eq:threeptdelproof}
	\begin{split}
		&\int d^{4}y \, d^{4}k \, d^{4}x \, \dfrac{e^{i(\omega_1 \Tilde{q}_1 + \omega_2 \Tilde{q}_2 + k)\cdot x} e^{-i(\omega_3 \Tilde{q}_3 + k)\cdot y}}{k^2 + m^2 + i\epsilon} \\
		& ~ = \int d^{4}y \, d^{4}k \, \dfrac{(2\pi)^{4}\delta^{(4)}(\omega_1 \Tilde{q}_1 + \omega_2 \Tilde{q}_2 + k)e^{-i(\omega_3 \Tilde{q}_3 +k)\cdot y}}{k^2+m^2+i\epsilon} \\
		& ~ = \int d^{4}y \, \dfrac{e^{-i(\omega_3 \Tilde{q}_3-\omega_1 \Tilde{q}_1 - \omega_2 \Tilde{q}_2)\cdot y}}{(\omega_1 \Tilde{q}_1+\omega_2 \Tilde{q}_2)^2 + m^2 + i\epsilon} \\
		& ~ = (2\pi)^{8}\dfrac{\delta^{(4)}(\omega_3 \Tilde{q}_3 - \omega_1 \Tilde{q}_1 - \omega_2 \Tilde{q}_2)}{2\omega_1\omega_2 \Tilde{q}_1\cdot \Tilde{q}_2 + m^2 + i\epsilon} \, 
	\end{split}
\end{equation}
which proves eq.\eqref{eq:threepointdelta}.

\subsubsection{Collinear limit}
\label{ap:threepointcollinear}

In this section, we will prove the result of three point function in the collinear limit given by eq.\eqref{eq:threepointcollinear}. We start with the momentum conserving delta function of eq.\eqref{eq:momcons2}
\begin{equation}
	\delta^{4}(\omega_3 \Tilde{q}_3 -\omega_1\tilde{q}_1 -\omega_2\Tilde{q}_2) = \dfrac{1}{\omega^3_3}\delta(\omega_3 - \omega_1 - \omega_2)\delta(z_{12})\delta(z_{13})\delta(\bar{z}_{12})\delta(\bar{z}_{13}) \, .
\end{equation}
This particular split of the delta function is lorentz invariant via the analysis of Appendix \ref{ap:lorentzdelta}. Since $\Tilde{q}_1 = (1,\Omega_1)$ is given by eq.\eqref{eq:4dparamq}, we see that $\Tilde{q}_1 = \Tilde{q}_2 = \tilde{q}_3$ corresponds to the collinear case when $\Omega_1 = \Omega_2 = \Omega_3$. For this case, we have from eq.\eqref{eq:threeptdelproof} in the analysis of Appendix \ref{ap:threepointcalc},
\begin{equation}
	\begin{split}
		&\int d^{4}y \, d^{4}k \, d^{4}x \, \dfrac{e^{i(\omega_1 \Tilde{q}_1 + \omega_2 \Tilde{q}_2 + k)\cdot x} e^{-i(\omega_3 \Tilde{q}_3 + k)\cdot y}}{k^2 + m^2 + i\epsilon} \\
		& ~ = (2\pi)^{8}\dfrac{\delta^{(4)}(\omega_3 \Tilde{q}_3 - \omega_1 \tilde{q}_1 - \omega_2 \Tilde{q}_2)}{ m^2 + i\epsilon} \, \\
		& ~ = (2\pi)^{8}\dfrac{\delta(\omega_3 - \omega_1 -\omega_2)\delta^2(z_{12})\delta^3(z_{13})}{ \omega^3_3 \, m^2 + i\epsilon} \, .
	\end{split}
\end{equation}
In the last step, one should keep in mind the subtle role played by the antipodal matching condition of eq.\eqref{eq:antipodal} elaborated in Appendix \ref{ap:twopointcalc}. Substituting the above equation in eq.\eqref{eq:threepointflatsecond}, we find 
\begin{equation}\label{eq:threepointflatthird}
	\begin{split}
		\langle O_{\Delta_1}(\p_1)O_{\Delta_2}(\p_2)O_{\Delta_3}(\p_3) \rangle \simeq 
		\mathcal{A}_{(3)} \int d\omega_1 d\omega_2 d\omega_3 \, \omega^{\Delta_1 -1}_1 \omega^{\Delta_2-1}_2 \omega^{\Delta_3-4}_3 \\
		\times e^{i\omega_1 u_1}e^{i\omega_2 u_2} e^{-i\omega_3 u_3}e^{-\epsilon(\omega_1+\omega_2+\omega_3)}\delta(\omega_3 -\omega_1 -\omega_2)\delta^2(z_{12})\delta^2(z_{13}) \, ,
	\end{split}
\end{equation}
where 
\begin{equation}
	\mathcal{A}_{(3)} = \dfrac{(2\pi)^{4}(i\mu)N^3_{\Delta_1}N^3_{\Delta_2}N^3_{\Delta_3}}{m^2 + i\epsilon} \, .
\end{equation}
We can perform the integral over $\omega_3$ using the delta function in eq.\eqref{eq:threepointflatthird} to obtain,
\begin{equation}
	\begin{split}
		& = \mathcal{A}_{(3)} \delta^2(z_{12})\delta^2(z_{13}) \int d\omega_1 d\omega_2 \, \omega^{\Delta_1 - 1}_1 \omega^{\Delta_2-1}_2 (\omega_1+\omega_2)^{\Delta_3-4} e^{i\omega_1(u_1-u_3)}e^{i\omega_2(u_2-u_3)}e^{-2\epsilon(\omega_1+\omega_2)} \\
		& = \mathcal{A}_{(3)} \delta^2(z_{12})\delta^2(z_{13}) \int d\omega_1 d\omega_2 \, \omega^{\Delta_1 - 1}_1 \omega^{\Delta_2-1}_2 \\
		& \hspace{4.5cm}\times  \sum_{k=0}^{\Delta_3 -4} {}^{\Delta_3-4}C_k \, \omega^k_1 \omega^{\Delta_3-4-k}_2 e^{i\omega_1(u_1-u_3)}e^{i\omega_2(u_2-u_3)} e^{-2\epsilon(\omega_1+\omega_2)} \\
		& = \mathcal{A}_{(3)} \delta^2(z_{12})\delta^2(z_{13}) \sum_{k=0}^{\Delta_3 -4} {}^{\Delta_3-4}C_k \int d\omega_1 \, \omega^{\Delta_1+k-1}_1 e^{i\omega_1(u_1-u_3)} e^{-2\epsilon\omega_1} \\
		& \hspace{5cm} \times \int d\omega_2 \, \omega^{\Delta_2+\Delta_3 - 5 -k}_2 e^{i\omega_2(u_2-u_3)} e^{-2\epsilon \omega_2} \\
		& = \mathcal{A}_{(3)}\delta^2(z_{12})\delta^2(z_{13}) \sum_{k=0}^{\Delta_3 -4} \dfrac{{}^{\Delta_3-4}C_k \, \Gamma(k+\Delta_1) \, \Gamma(\Delta_2+\Delta_3-k-4)}{(i(u_3-u_1))^{\Delta_1+k} \, (i(u_3-u_2))^{\Delta_2+\Delta_3-4-k}} \, .
	\end{split}
\end{equation}
In the second step, we have used the binomial expansion and in the final step, we have used the formula eq.\eqref{eq:omegaint} (the $i\epsilon$ prescription should appropriately justify the formula). The result is valid for $\Delta_3 \in \mathbb{N}$ and $\Delta_3 \geq 4$. This proves eq.\eqref{eq:threepointcollinear}.

\subsection{Lorentz invariance of delta functions}
\label{ap:lorentzdelta}

In this section, we will check that the split of the momentum conserving delta functions given in eq.\eqref{eq:twopointdelta} and eq.\eqref{eq:momcons2} will preserve lorentz invariance. In this process, one must ensure that the lorentz symmetry is preserved. In $3+1$ dimensions, $SO(3,1) \equiv SL(2,\mathbb{C})$. Thus, one can use the action of $SL(2,\mathbb{C})$ on the celestial sphere to check the lorentz symmetry of the delta function. Under $SL(2,\mathbb{C})$,
\begin{equation}\label{eq:sl2ctransf}
	z \to \dfrac{az+b}{cz+d} \hspace{0.5cm}, ~~ ad-bc=1
\end{equation}
Thus,
\begin{equation}\label{eq:zijsl2c}
	z_{ij} = z_i - z_j \to \dfrac{z_{ij}}{(cz_i+d)(cz_j+d)} \, .
\end{equation}
To find the transformation of the energies $\omega_i$ under $SL(2,\mathbb{C})$, we must look at the Mandelstam invariants $s_{ij} = (p_i+p_j)^2$, where $p^{\mu}_i = \omega_i \tilde{q}^{\mu}_i$ with $\tilde{q}^{\mu}_i$ given by the parametrization of eq.\eqref{eq:4dparamq}. Lorentz invariance of the Mandelstam invariants would require
\begin{equation}\label{eq:omegaisl2c}
	\omega_i \to (cz_i+d)(c\bar{z}_i+d) \,\omega_i ~~~ \text{under} ~~~ SL(2,\mathbb{C}) \, .
\end{equation}

Let us check this for eq.\eqref{eq:twopointdelta}:
\begin{equation}
	\delta^{(4)}(\omega_2 \tilde{q}^{\mu}_2-\omega_1 \tilde{q}^{\mu}_1) = \dfrac{1}{\omega_1}\delta(\omega_2-\omega_1)\delta(z_2-z_1)\delta(\bar{z}_2-\bar{z}_1) \, .
\end{equation}
$\delta^2(z_{12})$ will set $z_1 =z_2$,$\bar{z}_1=\bar{z}_2$. Thus, under an $SL(2,\mathbb{C})$ transformation given by eq.\eqref{eq:sl2ctransf}, we have the following
\begin{equation}
	\begin{split}
		\delta(\omega_2-\omega_1) &\to \delta \left((c^2 z_1 \bar{z}_1 + dc\bar{z}_1 + cd z_1 + d^2)(\omega_2-\omega_1) \right) = \dfrac{\delta(\omega_2-\omega_1)}{(cz_1+d)(c\bar{z}_1+d)} \\
		\delta(z_{12}) &\to \delta \left(\dfrac{z_{12}}{(cz_1+d)(cz_2+d)} \right) = (cz_1+d)^2 \delta(z_{12}) \\
		\delta(\bar{z}_{12}) &\to \delta \left(\dfrac{\bar{z}_{12}}{(c\bar{z}_1+d)(c\bar{z}_2+d)} \right) = (c\bar{z}_1+d)^2 \delta(\bar{z}_{12}) \\
		\dfrac{1}{\omega_1} &\to \dfrac{1}{\omega_1}\dfrac{1}{(cz_1+d)(c\bar{z}_1+d)}
	\end{split}
\end{equation}
Multiplying the terms in the RHS, we find that eq.\eqref{eq:twopointdelta} is lorentz invariant.

Now let us check the lorentz invariance of eq.\eqref{eq:momcons2}:
\begin{equation}
	\delta^{4}(\omega_3 \Tilde{q}_3 -\omega_1\tilde{q}_1 -\omega_2\Tilde{q}_2) = \dfrac{1}{\omega^3_3}\delta(\omega_3 - \omega_1 - \omega_2)\delta(z_{12})\delta(z_{13})\delta(\bar{z}_{12})\delta(\bar{z}_{13}) \, .
\end{equation}
To make the analysis simple, let us use a particular non-trivial case of the $SL(2,\mathbb{C})$ transformation of eq.\eqref{eq:sl2ctransf} given by $a=0,b=1,c=-1,d=0$:
\begin{equation}
	z \to -\dfrac{1}{z} \, .
\end{equation}
Thus, we have
\begin{equation}
	z_{ij} \to \dfrac{z_{ij}}{z_i z_j} ~~~, ~~~ \bar{z}_{ij} \to \dfrac{\bar{z}_{ij}}{\bar{z}_i\bar{z}_j} ~~~, ~~~ \omega_i \to z_i \bar{z}_i \omega_i 
\end{equation}
This implies
\begin{equation}
	\begin{split}
		\delta(\omega_3-\omega_1-\omega_2) &\to \delta(z_1\bar{z}_1 (\omega_3 - \omega_1 - \omega_2) ) = \dfrac{1}{z_1\bar{z}_1}\delta(\omega_3-\omega_1-\omega_2) \\
		\delta(z_{ij}) &\to \delta\left(\dfrac{z_{ij}}{z_iz_j}  \right) = z^2_1 \delta(z_{ij}) \\
		\delta(\bar{z}_{ij}) &\to \delta\left(\dfrac{\bar{z}_{ij}}{\bar{z}_i \bar{z}_j}  \right) = \bar{z}^2_1 \delta(\bar{z}_{ij}) \\
		\dfrac{1}{\omega^3_3} &\to \dfrac{1}{\omega^3_3}\dfrac{1}{(z_1\bar{z}_1)^3}
	\end{split}
\end{equation}
The extra factors are
\begin{equation}
	\dfrac{1}{(z_1\bar{z}_1)^3}\dfrac{1}{z_1\bar{z}_1} \, z^2_1 \, z^2_1 \, \bar{z}^2_1 \, \bar{z}^2_1 = 1 \, . 
\end{equation}
This proves that the delta function split in eq.\eqref{eq:momcons2} is Lorentz invariant.

\newpage

\bibliographystyle{JHEP}
\bibliography{flat}

\providecommand{\href}[2]{#2}\begingroup\raggedright\begin{thebibliography}{10}

\bibitem{tHooft:1993dmi}
G.~'t~Hooft, \emph{{Dimensional reduction in quantum gravity}}, {\emph{Conf.
  Proc. C} {\bfseries 930308} (1993) 284}
  [\href{https://arxiv.org/abs/gr-qc/9310026}{{\ttfamily gr-qc/9310026}}].

\bibitem{Susskind:1994vu}
L.~Susskind, \emph{{The World as a hologram}},
  \href{https://doi.org/10.1063/1.531249}{\emph{J. Math. Phys.} {\bfseries 36}
  (1995) 6377} [\href{https://arxiv.org/abs/hep-th/9409089}{{\ttfamily
  hep-th/9409089}}].

\bibitem{Maldacena:1997re}
J.~M. Maldacena, \emph{{The Large N limit of superconformal field theories and
  supergravity}}, \href{https://doi.org/10.1023/A:1026654312961}{\emph{Adv.
  Theor. Math. Phys.} {\bfseries 2} (1998) 231}
  [\href{https://arxiv.org/abs/hep-th/9711200}{{\ttfamily hep-th/9711200}}].

\bibitem{Witten:1998qj}
E.~Witten, \emph{{Anti-de Sitter space and holography}},
  \href{https://doi.org/10.4310/ATMP.1998.v2.n2.a2}{\emph{Adv. Theor. Math.
  Phys.} {\bfseries 2} (1998) 253}
  [\href{https://arxiv.org/abs/hep-th/9802150}{{\ttfamily hep-th/9802150}}].

\bibitem{Strominger:2013jfa}
A.~Strominger, \emph{{On BMS Invariance of Gravitational Scattering}},
  \href{https://doi.org/10.1007/JHEP07(2014)152}{\emph{JHEP} {\bfseries 07}
  (2014) 152} [\href{https://arxiv.org/abs/1312.2229}{{\ttfamily 1312.2229}}].

\bibitem{He:2014laa}
T.~He, V.~Lysov, P.~Mitra and A.~Strominger, \emph{{BMS supertranslations and
  Weinberg\textquoteright{}s soft graviton theorem}},
  \href{https://doi.org/10.1007/JHEP05(2015)151}{\emph{JHEP} {\bfseries 05}
  (2015) 151} [\href{https://arxiv.org/abs/1401.7026}{{\ttfamily 1401.7026}}].

\bibitem{Strominger:2014pwa}
A.~Strominger and A.~Zhiboedov, \emph{{Gravitational Memory, BMS
  Supertranslations and Soft Theorems}},
  \href{https://doi.org/10.1007/JHEP01(2016)086}{\emph{JHEP} {\bfseries 01}
  (2016) 086} [\href{https://arxiv.org/abs/1411.5745}{{\ttfamily 1411.5745}}].

\bibitem{Strominger:2017zoo}
A.~Strominger, \emph{{Lectures on the Infrared Structure of Gravity and Gauge
  Theory}},  \href{https://arxiv.org/abs/1703.05448}{{\ttfamily 1703.05448}}.

\bibitem{Pasterski:2021rjz}
S.~Pasterski, \emph{{Lectures on celestial amplitudes}},
  \href{https://doi.org/10.1140/epjc/s10052-021-09846-7}{\emph{Eur. Phys. J. C}
  {\bfseries 81} (2021) 1062}
  [\href{https://arxiv.org/abs/2108.04801}{{\ttfamily 2108.04801}}].

\bibitem{Raclariu:2021zjz}
A.-M. Raclariu, \emph{{Lectures on Celestial Holography}},
  \href{https://arxiv.org/abs/2107.02075}{{\ttfamily 2107.02075}}.

\bibitem{Bagchi:2016bcd}
A.~Bagchi, R.~Basu, A.~Kakkar and A.~Mehra, \emph{{Flat Holography: Aspects of
  the dual field theory}},
  \href{https://doi.org/10.1007/JHEP12(2016)147}{\emph{JHEP} {\bfseries 12}
  (2016) 147} [\href{https://arxiv.org/abs/1609.06203}{{\ttfamily
  1609.06203}}].

\bibitem{Bondi:1962px}
H.~Bondi, M.~G.~J. van~der Burg and A.~W.~K. Metzner, \emph{{Gravitational
  waves in general relativity. 7. Waves from axisymmetric isolated systems}},
  \href{https://doi.org/10.1098/rspa.1962.0161}{\emph{Proc. Roy. Soc. Lond. A}
  {\bfseries 269} (1962) 21}.

\bibitem{Sachs:1962zza}
R.~Sachs, \emph{{Asymptotic symmetries in gravitational theory}},
  \href{https://doi.org/10.1103/PhysRev.128.2851}{\emph{Phys. Rev.} {\bfseries
  128} (1962) 2851}.

\bibitem{Bagchi:2010zz}
A.~Bagchi, \emph{{Correspondence between Asymptotically Flat Spacetimes and
  Nonrelativistic Conformal Field Theories}},
  \href{https://doi.org/10.1103/PhysRevLett.105.171601}{\emph{Phys. Rev. Lett.}
  {\bfseries 105} (2010) 171601}
  [\href{https://arxiv.org/abs/1006.3354}{{\ttfamily 1006.3354}}].

\bibitem{Bagchi:2012cy}
A.~Bagchi and R.~Fareghbal, \emph{{BMS/GCA Redux: Towards Flatspace Holography
  from Non-Relativistic Symmetries}},
  \href{https://doi.org/10.1007/JHEP10(2012)092}{\emph{JHEP} {\bfseries 10}
  (2012) 092} [\href{https://arxiv.org/abs/1203.5795}{{\ttfamily 1203.5795}}].

\bibitem{Bagchi:2012xr}
A.~Bagchi, S.~Detournay, R.~Fareghbal and J.~Sim\'on, \emph{{Holography of 3D
  Flat Cosmological Horizons}},
  \href{https://doi.org/10.1103/PhysRevLett.110.141302}{\emph{Phys. Rev. Lett.}
  {\bfseries 110} (2013) 141302}
  [\href{https://arxiv.org/abs/1208.4372}{{\ttfamily 1208.4372}}].

\bibitem{Barnich:2012xq}
G.~Barnich, \emph{{Entropy of three-dimensional asymptotically flat
  cosmological solutions}},
  \href{https://doi.org/10.1007/JHEP10(2012)095}{\emph{JHEP} {\bfseries 10}
  (2012) 095} [\href{https://arxiv.org/abs/1208.4371}{{\ttfamily 1208.4371}}].

\bibitem{Bagchi:2015wna}
A.~Bagchi, D.~Grumiller and W.~Merbis, \emph{{Stress tensor correlators in
  three-dimensional gravity}},
  \href{https://doi.org/10.1103/PhysRevD.93.061502}{\emph{Phys. Rev. D}
  {\bfseries 93} (2016) 061502}
  [\href{https://arxiv.org/abs/1507.05620}{{\ttfamily 1507.05620}}].

\bibitem{Bagchi:2014iea}
A.~Bagchi, R.~Basu, D.~Grumiller and M.~Riegler, \emph{{Entanglement entropy in
  Galilean conformal field theories and flat holography}},
  \href{https://doi.org/10.1103/PhysRevLett.114.111602}{\emph{Phys. Rev. Lett.}
  {\bfseries 114} (2015) 111602}
  [\href{https://arxiv.org/abs/1410.4089}{{\ttfamily 1410.4089}}].

\bibitem{Jiang:2017ecm}
H.~Jiang, W.~Song and Q.~Wen, \emph{{Entanglement Entropy in Flat Holography}},
  \href{https://doi.org/10.1007/JHEP07(2017)142}{\emph{JHEP} {\bfseries 07}
  (2017) 142} [\href{https://arxiv.org/abs/1706.07552}{{\ttfamily
  1706.07552}}].

\bibitem{Hijano:2017eii}
E.~Hijano and C.~Rabideau, \emph{{Holographic entanglement and Poincar\'e
  blocks in three-dimensional flat space}},
  \href{https://doi.org/10.1007/JHEP05(2018)068}{\emph{JHEP} {\bfseries 05}
  (2018) 068} [\href{https://arxiv.org/abs/1712.07131}{{\ttfamily
  1712.07131}}].

\bibitem{Barnich:2015mui}
G.~Barnich, H.~A. Gonzalez, A.~Maloney and B.~Oblak, \emph{{One-loop partition
  function of three-dimensional flat gravity}},
  \href{https://doi.org/10.1007/JHEP04(2015)178}{\emph{JHEP} {\bfseries 04}
  (2015) 178} [\href{https://arxiv.org/abs/1502.06185}{{\ttfamily
  1502.06185}}].

\bibitem{Bagchi:2020rwb}
A.~Bagchi, P.~Nandi, A.~Saha and Zodinmawia, \emph{{BMS Modular Diaries: Torus
  one-point function}},
  \href{https://doi.org/10.1007/JHEP11(2020)065}{\emph{JHEP} {\bfseries 11}
  (2020) 065} [\href{https://arxiv.org/abs/2007.11713}{{\ttfamily
  2007.11713}}].

\bibitem{Bagchi:2012yk}
A.~Bagchi, S.~Detournay and D.~Grumiller, \emph{{Flat-Space Chiral Gravity}},
  \href{https://doi.org/10.1103/PhysRevLett.109.151301}{\emph{Phys. Rev. Lett.}
  {\bfseries 109} (2012) 151301}
  [\href{https://arxiv.org/abs/1208.1658}{{\ttfamily 1208.1658}}].

\bibitem{Barnich:2012aw}
G.~Barnich, A.~Gomberoff and H.~A. Gonzalez, \emph{{The Flat limit of three
  dimensional asymptotically anti-de Sitter spacetimes}},
  \href{https://doi.org/10.1103/PhysRevD.86.024020}{\emph{Phys. Rev. D}
  {\bfseries 86} (2012) 024020}
  [\href{https://arxiv.org/abs/1204.3288}{{\ttfamily 1204.3288}}].

\bibitem{Hartong:2015usd}
J.~Hartong, \emph{{Holographic Reconstruction of 3D Flat Space-Time}},
  \href{https://doi.org/10.1007/JHEP10(2016)104}{\emph{JHEP} {\bfseries 10}
  (2016) 104} [\href{https://arxiv.org/abs/1511.01387}{{\ttfamily
  1511.01387}}].

\bibitem{Bagchi:2022emh}
A.~Bagchi, S.~Banerjee, R.~Basu and S.~Dutta, \emph{{Scattering Amplitudes:
  Celestial and Carrollian}},
  \href{https://doi.org/10.1103/PhysRevLett.128.241601}{\emph{Phys. Rev. Lett.}
  {\bfseries 128} (2022) 241601}
  [\href{https://arxiv.org/abs/2202.08438}{{\ttfamily 2202.08438}}].

\bibitem{Pasterski:2017kqt}
S.~Pasterski and S.-H. Shao, \emph{{Conformal basis for flat space
  amplitudes}}, \href{https://doi.org/10.1103/PhysRevD.96.065022}{\emph{Phys.
  Rev. D} {\bfseries 96} (2017) 065022}
  [\href{https://arxiv.org/abs/1705.01027}{{\ttfamily 1705.01027}}].

\bibitem{Pasterski:2016qvg}
S.~Pasterski, S.-H. Shao and A.~Strominger, \emph{{Flat Space Amplitudes and
  Conformal Symmetry of the Celestial Sphere}},
  \href{https://doi.org/10.1103/PhysRevD.96.065026}{\emph{Phys. Rev. D}
  {\bfseries 96} (2017) 065026}
  [\href{https://arxiv.org/abs/1701.00049}{{\ttfamily 1701.00049}}].

\bibitem{Banerjee:2018gce}
S.~Banerjee, \emph{{Null Infinity and Unitary Representation of The Poincare
  Group}}, \href{https://doi.org/10.1007/JHEP01(2019)205}{\emph{JHEP}
  {\bfseries 01} (2019) 205}
  [\href{https://arxiv.org/abs/1801.10171}{{\ttfamily 1801.10171}}].

\bibitem{Donnay:2022aba}
L.~Donnay, A.~Fiorucci, Y.~Herfray and R.~Ruzziconi, \emph{{Carrollian
  Perspective on Celestial Holography}},
  \href{https://doi.org/10.1103/PhysRevLett.129.071602}{\emph{Phys. Rev. Lett.}
  {\bfseries 129} (2022) 071602}
  [\href{https://arxiv.org/abs/2202.04702}{{\ttfamily 2202.04702}}].

\bibitem{Donnay:2022wvx}
L.~Donnay, A.~Fiorucci, Y.~Herfray and R.~Ruzziconi, \emph{{Bridging Carrollian
  and Celestial Holography}},
  \href{https://arxiv.org/abs/2212.12553}{{\ttfamily 2212.12553}}.

\bibitem{PipolodeGioia:2022exe}
L.~Pipolode~Gioia and A.-M. Raclariu, \emph{{Eikonal Approximation in Celestial
  CFT}},  \href{https://arxiv.org/abs/2206.10547}{{\ttfamily 2206.10547}}.

\bibitem{Duval:2014uva}
C.~Duval, G.~W. Gibbons and P.~A. Horvathy, \emph{{Conformal Carroll groups and
  BMS symmetry}},
  \href{https://doi.org/10.1088/0264-9381/31/9/092001}{\emph{Class. Quant.
  Grav.} {\bfseries 31} (2014) 092001}
  [\href{https://arxiv.org/abs/1402.5894}{{\ttfamily 1402.5894}}].

\bibitem{Bagchi:2022nvj}
A.~Bagchi, A.~Banerjee and H.~Muraki, \emph{{Boosting to BMS}},
  \href{https://doi.org/10.1007/JHEP09(2022)251}{\emph{JHEP} {\bfseries 09}
  (2022) 251} [\href{https://arxiv.org/abs/2205.05094}{{\ttfamily
  2205.05094}}].

\bibitem{Bagchi:2023ysc}
A.~Bagchi, K.~S. Kolekar and A.~Shukla, \emph{{Carrollian Origins of Bjorken
  Flow}},  \href{https://arxiv.org/abs/2302.03053}{{\ttfamily 2302.03053}}.

\bibitem{LevyLeblond}
L.~Leblond, \emph{{Une nouvelle limite non-relativiste du group de Poincaré}},
  {\emph{Annales Poincare Phys.Theor.} {\bfseries 3} (1965) }.

\bibitem{NDS}
N.~Sen~Gupta, \emph{{On an Analogue of the Galileo Group}}, {\emph{Nuovo Cim.
  54 (1966) 512 • DOI: 10.1007/BF02740871} }.

\bibitem{Henneaux:1979vn}
M.~Henneaux, \emph{{Geometry of Zero Signature Space-times}}, {\emph{Bull. Soc.
  Math. Belg.} {\bfseries 31} (1979) 47}.

\bibitem{Duval:2014uoa}
C.~Duval, G.~W. Gibbons, P.~A. Horvathy and P.~M. Zhang, \emph{{Carroll versus
  Newton and Galilei: two dual non-Einsteinian concepts of time}},
  \href{https://doi.org/10.1088/0264-9381/31/8/085016}{\emph{Class. Quant.
  Grav.} {\bfseries 31} (2014) 085016}
  [\href{https://arxiv.org/abs/1402.0657}{{\ttfamily 1402.0657}}].

\bibitem{Barnich:2010eb}
G.~Barnich and C.~Troessaert, \emph{{Aspects of the BMS/CFT correspondence}},
  \href{https://doi.org/10.1007/JHEP05(2010)062}{\emph{JHEP} {\bfseries 05}
  (2010) 062} [\href{https://arxiv.org/abs/1001.1541}{{\ttfamily 1001.1541}}].

\bibitem{Campiglia:2014yka}
M.~Campiglia and A.~Laddha, \emph{{Asymptotic symmetries and subleading soft
  graviton theorem}},
  \href{https://doi.org/10.1103/PhysRevD.90.124028}{\emph{Phys. Rev.}
  {\bfseries D90} (2014) 124028}
  [\href{https://arxiv.org/abs/1408.2228}{{\ttfamily 1408.2228}}].

\bibitem{Schwarz:2022dqf}
J.~H. Schwarz, \emph{{Diffeomorphism Symmetry in Two Dimensions and Celestial
  Holography}},  \href{https://arxiv.org/abs/2208.13304}{{\ttfamily
  2208.13304}}.

\bibitem{Dutta:2022vkg}
S.~Dutta, \emph{{Stress tensors of 3d Carroll CFTs}},
  \href{https://arxiv.org/abs/2212.11002}{{\ttfamily 2212.11002}}.

\bibitem{Banerjee:2020zlg}
S.~Banerjee, S.~Ghosh and P.~Paul, \emph{{MHV graviton scattering amplitudes
  and current algebra on the celestial sphere}},
  \href{https://doi.org/10.1007/JHEP02(2021)176}{\emph{JHEP} {\bfseries 02}
  (2021) 176} [\href{https://arxiv.org/abs/2008.04330}{{\ttfamily
  2008.04330}}].

\bibitem{Strominger:2021mtt}
A.~Strominger, \emph{{$w_{1+\infty}$ Algebra and the Celestial Sphere: Infinite
  Towers of Soft Graviton, Photon, and Gluon Symmetries}},
  \href{https://doi.org/10.1103/PhysRevLett.127.221601}{\emph{Phys. Rev. Lett.}
  {\bfseries 127} (2021) 221601}.

\bibitem{Chen:2021xkw}
B.~Chen, R.~Liu and Y.-f. Zheng, \emph{{On Higher-dimensional Carrollian and
  Galilean Conformal Field Theories}},
  \href{https://arxiv.org/abs/2112.10514}{{\ttfamily 2112.10514}}.

\bibitem{deBoer:2021jej}
J.~de~Boer, J.~Hartong, N.~A. Obers, W.~Sybesma and S.~Vandoren, \emph{{Carroll
  Symmetry, Dark Energy and Inflation}},
  \href{https://doi.org/10.3389/fphy.2022.810405}{\emph{Front. in Phys.}
  {\bfseries 10} (2022) 810405}
  [\href{https://arxiv.org/abs/2110.02319}{{\ttfamily 2110.02319}}].

\bibitem{Chang:2022seh}
C.-M. Chang and W.-J. Ma, \emph{{Missing Corner in the Sky: Massless
  Three-Point Celestial Amplitudes}},
  \href{https://arxiv.org/abs/2212.07025}{{\ttfamily 2212.07025}}.

\bibitem{Atanasov:2021oyu}
A.~Atanasov, A.~Ball, W.~Melton, A.-M. Raclariu and A.~Strominger, \emph{{(2,
  2) Scattering and the celestial torus}},
  \href{https://doi.org/10.1007/JHEP07(2021)083}{\emph{JHEP} {\bfseries 07}
  (2021) 083} [\href{https://arxiv.org/abs/2101.09591}{{\ttfamily
  2101.09591}}].

\bibitem{Polchinski:1999ry}
J.~Polchinski, \emph{{S matrices from AdS space-time}},
  \href{https://arxiv.org/abs/hep-th/9901076}{{\ttfamily hep-th/9901076}}.

\bibitem{Susskind:1998vk}
L.~Susskind, \emph{{Holography in the flat space limit}},
  \href{https://doi.org/10.1063/1.1301570}{\emph{AIP Conf. Proc.} {\bfseries
  493} (1999) 98} [\href{https://arxiv.org/abs/hep-th/9901079}{{\ttfamily
  hep-th/9901079}}].

\bibitem{Giddings:1999jq}
S.~B. Giddings, \emph{{Flat space scattering and bulk locality in the AdS / CFT
  correspondence}},
  \href{https://doi.org/10.1103/PhysRevD.61.106008}{\emph{Phys. Rev. D}
  {\bfseries 61} (2000) 106008}
  [\href{https://arxiv.org/abs/hep-th/9907129}{{\ttfamily hep-th/9907129}}].

\bibitem{Balasubramanian:1999ri}
V.~Balasubramanian, S.~B. Giddings and A.~E. Lawrence, \emph{{What do CFTs tell
  us about Anti-de Sitter space-times?}},
  \href{https://doi.org/10.1088/1126-6708/1999/03/001}{\emph{JHEP} {\bfseries
  03} (1999) 001} [\href{https://arxiv.org/abs/hep-th/9902052}{{\ttfamily
  hep-th/9902052}}].

\bibitem{Giddings:1999qu}
S.~B. Giddings, \emph{{The Boundary S matrix and the AdS to CFT dictionary}},
  \href{https://doi.org/10.1103/PhysRevLett.83.2707}{\emph{Phys. Rev. Lett.}
  {\bfseries 83} (1999) 2707}
  [\href{https://arxiv.org/abs/hep-th/9903048}{{\ttfamily hep-th/9903048}}].

\bibitem{PhysRevD.94.065017}
M.~Gary and S.~B. Giddings, \emph{Constraints on a fine-grained ads/cft
  correspondence},
  \href{https://doi.org/10.1103/PhysRevD.94.065017}{\emph{Phys. Rev. D}
  {\bfseries 94} (2016) 065017}.

\bibitem{Gary:2009ae}
M.~Gary, S.~B. Giddings and J.~Penedones, \emph{{Local bulk S-matrix elements
  and CFT singularities}},
  \href{https://doi.org/10.1103/PhysRevD.80.085005}{\emph{Phys. Rev. D}
  {\bfseries 80} (2009) 085005}
  [\href{https://arxiv.org/abs/0903.4437}{{\ttfamily 0903.4437}}].

\bibitem{Penedones:2010ue}
J.~Penedones, \emph{{Writing CFT correlation functions as AdS scattering
  amplitudes}}, \href{https://doi.org/10.1007/JHEP03(2011)025}{\emph{JHEP}
  {\bfseries 03} (2011) 025} [\href{https://arxiv.org/abs/1011.1485}{{\ttfamily
  1011.1485}}].

\bibitem{Mack:2009mi}
G.~Mack, \emph{{D-independent representation of Conformal Field Theories in D
  dimensions via transformation to auxiliary Dual Resonance Models. Scalar
  amplitudes}},  \href{https://arxiv.org/abs/0907.2407}{{\ttfamily 0907.2407}}.

\bibitem{Mack:2009gy}
G.~Mack, \emph{{D-dimensional Conformal Field Theories with anomalous
  dimensions as Dual Resonance Models}}, {\emph{Bulg. J. Phys.} {\bfseries 36}
  (2009) 214} [\href{https://arxiv.org/abs/0909.1024}{{\ttfamily 0909.1024}}].

\bibitem{Fitzpatrick:2011jn}
A.~L. Fitzpatrick and J.~Kaplan, \emph{{Scattering States in AdS/CFT}},
  \href{https://arxiv.org/abs/1104.2597}{{\ttfamily 1104.2597}}.

\bibitem{Fitzpatrick:2011hu}
A.~L. Fitzpatrick and J.~Kaplan, \emph{{Analyticity and the Holographic
  S-Matrix}}, \href{https://doi.org/10.1007/JHEP10(2012)127}{\emph{JHEP}
  {\bfseries 10} (2012) 127} [\href{https://arxiv.org/abs/1111.6972}{{\ttfamily
  1111.6972}}].

\bibitem{Fitzpatrick:2011dm}
A.~L. Fitzpatrick and J.~Kaplan, \emph{{Unitarity and the Holographic
  S-Matrix}}, \href{https://doi.org/10.1007/JHEP10(2012)032}{\emph{JHEP}
  {\bfseries 10} (2012) 032} [\href{https://arxiv.org/abs/1112.4845}{{\ttfamily
  1112.4845}}].

\bibitem{Paulos:2016fap}
M.~F. Paulos, J.~Penedones, J.~Toledo, B.~C. van Rees and P.~Vieira, \emph{{The
  S-matrix bootstrap. Part I: QFT in AdS}},
  \href{https://doi.org/10.1007/JHEP11(2017)133}{\emph{JHEP} {\bfseries 11}
  (2017) 133} [\href{https://arxiv.org/abs/1607.06109}{{\ttfamily
  1607.06109}}].

\bibitem{Raju:2012zr}
S.~Raju, \emph{{New Recursion Relations and a Flat Space Limit for AdS/CFT
  Correlators}}, \href{https://doi.org/10.1103/PhysRevD.85.126009}{\emph{Phys.
  Rev. D} {\bfseries 85} (2012) 126009}
  [\href{https://arxiv.org/abs/1201.6449}{{\ttfamily 1201.6449}}].

\bibitem{Hamilton:2005ju}
A.~Hamilton, D.~N. Kabat, G.~Lifschytz and D.~A. Lowe, \emph{{Local bulk
  operators in AdS/CFT: A Boundary view of horizons and locality}},
  \href{https://doi.org/10.1103/PhysRevD.73.086003}{\emph{Phys. Rev. D}
  {\bfseries 73} (2006) 086003}
  [\href{https://arxiv.org/abs/hep-th/0506118}{{\ttfamily hep-th/0506118}}].

\bibitem{Hamilton:2006az}
A.~Hamilton, D.~N. Kabat, G.~Lifschytz and D.~A. Lowe, \emph{{Holographic
  representation of local bulk operators}},
  \href{https://doi.org/10.1103/PhysRevD.74.066009}{\emph{Phys. Rev. D}
  {\bfseries 74} (2006) 066009}
  [\href{https://arxiv.org/abs/hep-th/0606141}{{\ttfamily hep-th/0606141}}].

\bibitem{Hamilton:2006fh}
A.~Hamilton, D.~N. Kabat, G.~Lifschytz and D.~A. Lowe, \emph{{Local bulk
  operators in AdS/CFT: A Holographic description of the black hole interior}},
  \href{https://doi.org/10.1103/PhysRevD.75.106001}{\emph{Phys. Rev. D}
  {\bfseries 75} (2007) 106001}
  [\href{https://arxiv.org/abs/hep-th/0612053}{{\ttfamily hep-th/0612053}}].

\bibitem{Hijano:2019qmi}
E.~Hijano, \emph{{Flat space physics from AdS/CFT}},
  \href{https://doi.org/10.1007/JHEP07(2019)132}{\emph{JHEP} {\bfseries 07}
  (2019) 132} [\href{https://arxiv.org/abs/1905.02729}{{\ttfamily
  1905.02729}}].

\bibitem{Hijano:2020szl}
E.~Hijano and D.~Neuenfeld, \emph{{Soft photon theorems from CFT Ward identites
  in the flat limit of AdS/CFT}},
  \href{https://doi.org/10.1007/JHEP11(2020)009}{\emph{JHEP} {\bfseries 11}
  (2020) 009} [\href{https://arxiv.org/abs/2005.03667}{{\ttfamily
  2005.03667}}].

\bibitem{Li:2021snj}
Y.-Z. Li, \emph{{Notes on flat-space limit of AdS/CFT}},
  \href{https://doi.org/10.1007/JHEP09(2021)027}{\emph{JHEP} {\bfseries 09}
  (2021) 027} [\href{https://arxiv.org/abs/2106.04606}{{\ttfamily
  2106.04606}}].

\bibitem{Casali:2022fro}
E.~Casali, W.~Melton and A.~Strominger, \emph{{Celestial amplitudes as
  AdS-Witten diagrams}},
  \href{https://doi.org/10.1007/JHEP11(2022)140}{\emph{JHEP} {\bfseries 11}
  (2022) 140} [\href{https://arxiv.org/abs/2204.10249}{{\ttfamily
  2204.10249}}].

\bibitem{Iacobacci:2022yjo}
L.~Iacobacci, C.~Sleight and M.~Taronna, \emph{{From Celestial Correlators to
  AdS, and back}},  \href{https://arxiv.org/abs/2208.01629}{{\ttfamily
  2208.01629}}.

\bibitem{Sleight:2023ojm}
C.~Sleight and M.~Taronna, \emph{{Celestial Holography Revisited}},
  \href{https://arxiv.org/abs/2301.01810}{{\ttfamily 2301.01810}}.

\bibitem{Lam:2017ofc}
H.~T. Lam and S.-H. Shao, \emph{{Conformal Basis, Optical Theorem, and the Bulk
  Point Singularity}},
  \href{https://doi.org/10.1103/PhysRevD.98.025020}{\emph{Phys. Rev. D}
  {\bfseries 98} (2018) 025020}
  [\href{https://arxiv.org/abs/1711.06138}{{\ttfamily 1711.06138}}].

\bibitem{Dirac:1936fq}
P.~A.~M. Dirac, \emph{{Wave equations in conformal space}},
  \href{https://doi.org/10.2307/1968455}{\emph{Annals Math.} {\bfseries 37}
  (1936) 429}.

\bibitem{Penedones:2007ns}
J.~Penedones, \emph{{High Energy Scattering in the AdS/CFT Correspondence}},
  other thesis, 12, 2007.

\bibitem{Bagchi:2016geg}
A.~Bagchi, M.~Gary and Zodinmawia, \emph{{Bondi-Metzner-Sachs bootstrap}},
  \href{https://doi.org/10.1103/PhysRevD.96.025007}{\emph{Phys. Rev. D}
  {\bfseries 96} (2017) 025007}
  [\href{https://arxiv.org/abs/1612.01730}{{\ttfamily 1612.01730}}].

\bibitem{Bagchi:2017cpu}
A.~Bagchi, M.~Gary and Zodinmawia, \emph{{The nuts and bolts of the BMS
  Bootstrap}}, \href{https://doi.org/10.1088/1361-6382/aa8003}{\emph{Class.
  Quant. Grav.} {\bfseries 34} (2017) 174002}
  [\href{https://arxiv.org/abs/1705.05890}{{\ttfamily 1705.05890}}].

\bibitem{Banerjee:2019prz}
S.~Banerjee, S.~Ghosh, P.~Pandey and A.~P. Saha, \emph{{Modified celestial
  amplitude in Einstein gravity}},
  \href{https://doi.org/10.1007/JHEP03(2020)125}{\emph{JHEP} {\bfseries 03}
  (2020) 125} [\href{https://arxiv.org/abs/1909.03075}{{\ttfamily
  1909.03075}}].

\bibitem{BURGESS1985137}
C.~Burgess and C.~Lütken, \emph{Propagators and effective potentials in
  anti-de sitter space},
  \href{https://doi.org/https://doi.org/10.1016/0370-2693(85)91415-7}{\emph{Physics
  Letters B} {\bfseries 153} (1985) 137}.

\end{thebibliography}\endgroup

\end{document}